\documentclass[twocolumn,showpacs,superscriptaddress,preprintnumbers,amsmath,amssymb,aps,prx]{revtex4-1}
\usepackage{graphicx}
\usepackage{mathptmx}
\usepackage{verbatim}
\usepackage{graphicx}
\usepackage{amsmath}
\usepackage{amssymb}
\usepackage{amsfonts}
\usepackage{mathrsfs}
\usepackage{amsmath} 
\usepackage{color}
\usepackage{graphicx}
\usepackage{bm} 
\usepackage{amssymb}
\usepackage{xspace}
\usepackage{hyperref}
\usepackage{multirow}
\usepackage{color}
\usepackage{float}
\usepackage{bbold}
\usepackage{tabu}
\usepackage{textcomp}
\usepackage{dcolumn}
\usepackage{appendix} 

\newcolumntype{L}[1]{>{\raggedright\arraybackslash}m{#1}}
\newcolumntype{C}[1]{>{\centering\arraybackslash}m{#1}}
\newcolumntype{R}[1]{>{\raggedleft\arraybackslash}m{#1}}
\newcolumntype{N}{@{}m{0pt}@{}}

\begin{document}
\title{
{\it Ab initio} tight-binding Models for Mono- and Bilayer Hexagonal Boron Nitride ({\it h}-BN)
}
\author{Srivani Javvaji}
\affiliation{Department of Physics, University of Seoul, Seoul 02504, Korea}
\author{Fengping Li}
\affiliation{Department of Physics, University of Seoul, Seoul 02504, Korea}
\author{Jeil Jung}
\email{jeiljung@uos.ac.kr}
\affiliation{Department of Physics, University of Seoul, Seoul 02504, Korea}
\affiliation{Department of Smart Cities, University of Seoul, Seoul 02504, Korea}

\begin{abstract}

Hexagonal boron nitride ({\it h}-BN) exhibits dominant $\pi$-bands near the Fermi level, similar to graphene. However, unlike graphene, where tight-binding (TB) models accurately reproduce band edges near the $K$ and $K^{\prime}$ points in the Brillouin zone, a wider bandgap in {\it h}-BN necessitates capturing the band edges at both the $K$ and $M$ points for precise bandgap calculations. We present effective TB models derived from {\it ab initio} calculations using maximally localized Wannier functions (MLWFs) centered on boron and nitrogen sites. These models consider hopping terms of up to four distant neighbors and achieve excellent agreement with {\it ab initio} results near the $K$ and $M$ points. Furthermore, we compare the band structures from our simplified models with those obtained from \textit{ab initio} calculations and the full tight-binding model to assess their accuracy. 
To account for the effects of strains, we introduce fitting parametrizations that relate the hopping parameters of the effective TB model to the lattice constant and interlayer distance. Additionally, we utilize the two-center approximation to calculate the interlayer hopping energies based on the relative distances between sublattices to generalize the interlayer hopping parameters across different stacking configurations. 
We demonstrate the effectiveness of this method by comparing the electronic structure of zero-twist and twisted {\it h}-BN systems with {\it ab initio} calculations.
\end{abstract}
%

%
%
\maketitle
\section{Introduction}  

Using {\it h}-BN as a two-dimensional substrate for graphene systems is prevalent in emerging electronic devices~\cite{substrate_expt}, moreover the resulting lattice mismatch and misorientation create moiré patterns that influence electronic properties~\cite{jj_PRB2014,jj_NatCom2015,GBN_Moire_expt}. 
Interestingly, twisted {\it h}-BN bilayers themselves exhibit unique electronic properties like flat bands, excitons, and strong correlations, making them intriguing systems, particularly due to their inherent bandgap~\cite{t2BN_2019,t2BN_2020, t2BN_2023}.
Similar to graphene, \textit{h}-BN features are dominated by $\pi$-bands near the Fermi level, which arise from the $\pi$-electrons in the $p_z$ orbitals of boron and nitrogen atoms. The overlap of these $p_z$ orbitals is crucial for constructing TB models for both \textit{h}-BN monolayers and bilayers. However, existing TB parametrizations struggle to accurately reproduce the $\pi$-band dispersion across the entire Brillouin zone, particularly away from the $K$-point. While these models successfully capture the band edges near the $K$ and $K^{\prime}$ points using methods such as {\em ab initio}-fitted minimal TB models~\cite{double_hBN_2011, single_hBN_2016, double_hBN_2018, previous_study_on_h-BN_TBmodel, symmetry2022} or $K$/$K^{\prime}$-based continuum Hamiltonians~\cite{MassiveDirac_JSV}, they often fail to accurately replicate the band structure around the $M$ point and the dispersion throughout the Brillouin zone. This limitation complicates the precise estimation of the bandgap and its nature (direct or indirect), which is crucial for various device applications.

\begin{figure}
\centering 
\includegraphics[width=\columnwidth]{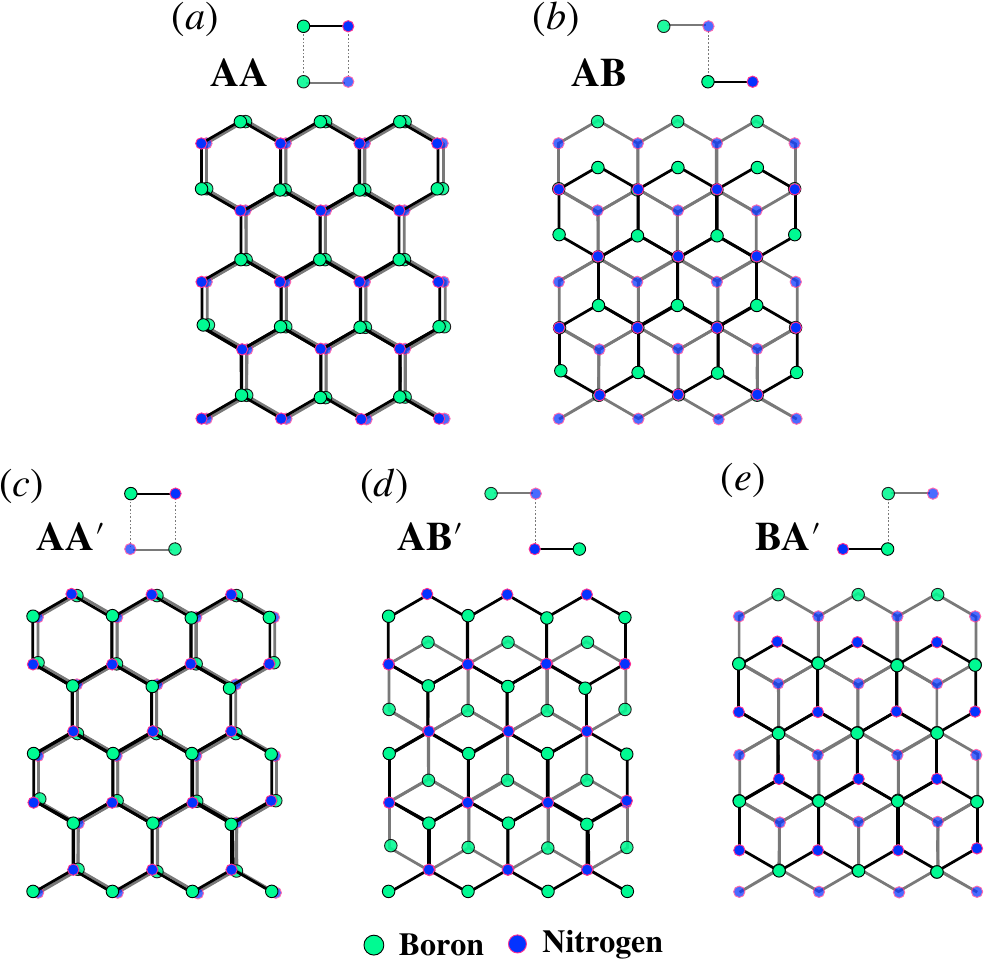} 
\caption{(Color online) Top views of the {\it h}-BN bilayers with standard stacking configurations: a) AA, b) AB, c) AA$^\prime$, d) AB$^\prime$ and e) BA$^\prime$. Green and blue circles represent boron and nitrogen atoms, respectively. Side views of each stacking is shown at the top of each sub-figure.  } 
\label{fig:1}
\end{figure}

In this paper, we propose simplified TB models for monolayer and bilayer \textit{h}-BN that consider only a limited number of distant neighbors. Despite the reduced complexity, our models accurately reproduce the band edges from density functional theory (DFT) calculations near high-symmetry points across the entire first Brillouin zone (FBZ). This allows for precise estimation of the bandgap and its characteristics in \textit{h}-BN systems. We provide the necessary hopping parameters for constructing the model Hamiltonians for the monolayer and bilayer configurations examined in this study, and demonstrate their effectiveness by comparing them with DFT band structure calculations. Additionally, we present effective hopping terms for modeling \textit{h}-BN bilayers under various conditions, with fitting parameters based on interlayer distance and lattice parameters. We also introduce a generalized two-center approximation to map interlayer hopping in both zero-twist and twisted bilayers, while using our models to accurately describe intra-layer interactions.

The manuscript is organized as follows: In Section II, we provide a brief summary of the $ab~initio $ calculations used in the study. In Section III, we present several TB model approximations to the first-principles-calculated band structures, and we provide the hopping parameters necessary to construct effective models. In Section IV, we present the revised two-center TB model for bilayer {\it h}-BN, and in Section V we present the conclusions of our findings.
\section{{\em Ab initio} calculation details}
\label{sec:methods}
We use the Quantum ESPRESSO package~\cite{QE_1,QE_2} for all our first-principles calculations, employing a plane-wave basis set~\cite{planewaves} within the local-density approximation (LDA) using the Perdew-Zunger parametrization~\cite{LDA}. For enhanced accuracy, we performed DFT calculations with a 42 × 42 × 1 $k$-point sampling density and a plane wave cutoff energy of 60 Ry. We constructed the \textit{h}-BN monolayer and bilayer structures using an LDA-DFT optimized lattice constant of $a = 2.48~\rm \AA$, which is slightly smaller than the experimental value of $2.504~\rm \AA$~\cite{Lynch}.

\begin{figure}[htb!]
\begin{center}
\includegraphics[width=6cm]{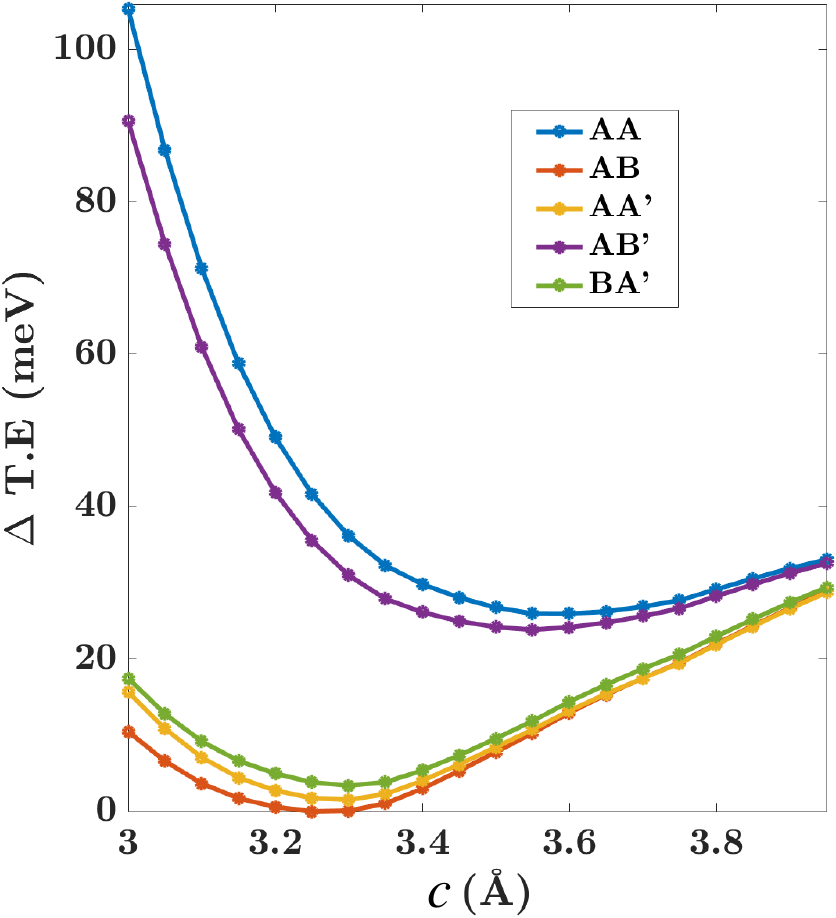}
\caption{(Color online) The total energy differences of various BLBN stackings are plotted against interlayer distance, relative to the most stable AB-stacking configuration at its equilibrium interlayer distance of \( c = 3.261~\text{Å} \). The calculations were performed using DFT within the LDA.} 
\label{fig:2}
\end{center}
\end{figure}

The {\it h}-BN bilayers can exist in five distinct stacking configurations~\cite{double_hBN_2011, double_hBN_2018, previous_study_on_h-BN_TBmodel, symmetry2022, 2010stacking, 2013stacking, SciPostPhys2023, Zettl}, as shown in Fig.~\ref{fig:1}. These stacking arrangements influence the equilibrium distance between the layers, as evident from the variation in total energy with interlayer spacing ($c$) in Fig.~\ref{fig:2}. It is noted that AB-stacking is energetically stable compared to rest of the configurations \cite{2010stacking, 2013stacking, SciPostPhys2023, Zettl}, followed by the AA$^\prime$-stacking with a total energy difference of 1.664 meV per unit cell, which benefits from attractive electrostatic interactions between vertically alternating boron and nitrogen atoms~\cite{2010stacking,2013stacking,SciPostPhys2023,Zettl}. Since the most stable AB-stacked bilayer has an equilibrium interlayer distance of \(c = 3.261 \, \text{Å}\) (Fig.~\ref{fig:2}), we use this distance for all other stacking configurations when constructing the TB models.
The TB Hamiltonian matrix elements are obtained by transforming the DFT Hamiltonian from the Bloch basis to the Wannier basis. This process involves first determining the electronic structure through DFT calculations. Subsequently, Wannier functions are generated from a set of localized orbitals centered at boron and nitrogen atoms using WANNIER90~\cite{MLWF}. The transformation is then facilitated by overlap matrices between Bloch states and Wannier functions, computed within the WANNIER90~\cite{MLWF} tool. This process directly provides the TB parameters: on-site energies from diagonal elements and hopping integrals from off-diagonal elements. This approach builds upon previous work developing a full tight-binding (FTB) model for both monolayer and bilayer graphene, which accurately reproduced low-energy bands from first-principles LDA calculations~\cite{previous_study_on_monoG_TBmodel, previous_study_on_BiG_TBmodel}.

\section{tight-binding models}
\label{sec:TB}
\subsection{Hamiltonian}
The Hamiltonian for the $\pi$-bands in BLBN can be represented by a $\bm k$ dependent 4×4 size matrix:

\begin{eqnarray}
H_{\rm BLBN} (\bm k) = \begin{pmatrix} H_{\rm MBN_1} (\bm k)   & H_{\rm Coup} (\bm k)   \\ H^{\dagger}_{\rm Coup}(\bm k)  &  H_{\rm MBN_2} (\bm k)   \end{pmatrix}
\label{Eq:Htbg}
\end{eqnarray}
where $H_{\rm MBN_1} (\bm k)$ and $ H_{\rm MBN_2} (\bm k)$ are $2\times2$ size Hamiltonian matrices that describe the inter- and intra-sublattice hopping processes within each layer. These matrices are given by:

\begin{eqnarray}
H_{\rm MBN_i} (\bm k) = \begin{pmatrix} H_{\rm B_iB_i} (\bm k)   & H_{\rm B_iN_i} (\bm k)   \\ H_{\rm N_iB_i} (\bm k)  &  H_{\rm N_iN_i} (\bm k)\end{pmatrix}    
\label{Eq:Htbg} 
\end{eqnarray}
where $i = 1,2$ denotes the indices for the top and bottom layers, respectively.
The coupling between these two layers is described by:

\begin{eqnarray}
H_{\rm Coup} (\bm k) = \begin{pmatrix} H_{\rm B_1B_2} (\bm k)   & H_{\rm B_1N_2} (\bm k)   \\ H_{\rm N_1B_2} (\bm k)  &  H_{\rm N_1N_2} (\bm k)   \end{pmatrix}
\label{Eq:Htbg}
\end{eqnarray}

Each element of the Hamiltonian matrix, representing hopping processes between a distinct pair of atomic orbitals, is defined as a sum over neighbor indices \( n \), as shown below:

\begin{equation}
\begin{aligned}
H_{\alpha\beta}(\bm{k}) & = \sum_n t_n~ e^{i \bm{k} \cdot \bm{R}_n}
\end{aligned}
\label{Eq:tb}
\end{equation}
Here, $t_n$ signifies the hopping energy, and $\bm R_n$ denotes the position vector describing the hopping process between sublattices $\alpha$ and $\beta$, which consist of boron or nitrogen atoms. The summation over the phase factor $e^{i \bm{k} \cdot \bm{R_n}}$ is known as the structure factor, accounting for the arrangement of the $n^{\rm th}$ neighbors~\cite{previous_study_on_monoG_TBmodel,previous_study_on_BiG_TBmodel}. This is denoted by $g_n(\bm k)$ and $f_n(\bm k)$ for the intra- and inter-sublattice hopping processes, respectively. For simplicity we will denote these structure factors as G$_n$ and F$_n$ in this paper.

The FTB model Hamiltonian is constructed by considering up to $n = 15$ neighbors for each element as described in Eq.(\ref{Eq:tb}). To reduce the complexity we build simpler effective TB models that only consider a limited number of nearest neighbors. These models are derived from the low-energy \(\bm{k}\cdot\bm{p}\) Hamiltonian, which is obtained by performing a Taylor expansion of the energy bands around the \(K\)-point~\cite{previous_study_on_monoG_TBmodel, previous_study_on_BiG_TBmodel}. In this \(\bm{k}\cdot\bm{p}\) framework, the Hamiltonian element \(H_{\alpha\beta}(\bm{k})\) is computed by expanding the bands at \(\bm{k} = \bm{k}_D + \bm{k}\), where \(\bm{k}_D\) is the \(K\)-point and \(\bm{k}\) is a small deviation vector. The details of this expansion are provided in Appendix~\ref{sec:appendixA}. This expansion decomposes $H_{\alpha\beta}(\bm{k})$ into two components, diagonal elements ($\alpha = \beta$) corresponding to intra-sublattice processes (see Eq.~\ref{Eq:Hamil-intra}) and off-diagonal elements ($\alpha \neq \beta$) representing inter-sublattice processes (see Eq.~\ref{Eq:Hamil-inter}). The zeroth-order expansion coefficient, $C'_{\alpha\beta0}$, governing intra-sublattice processes, and the first-order coefficient, $C_{\alpha\beta1}$, for inter-sublattice processes are essential for constructing the effective Hamiltonian. These coefficients are derived from the hopping parameters, as expressed in Eqs.\ref{Eq:cpab0} and \ref{Eq:cab1}~\cite{previous_study_on_monoG_TBmodel}.

\begin{figure}
\centering 
\includegraphics[width=\columnwidth]{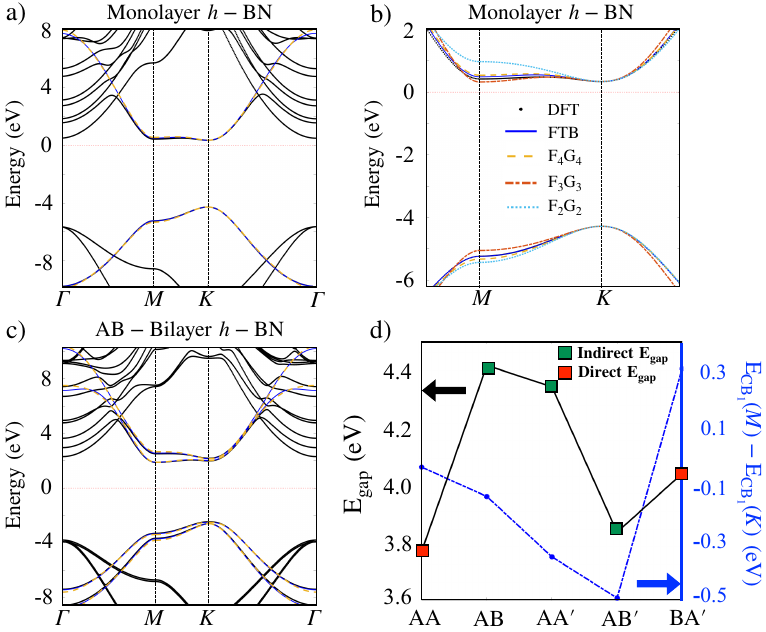}
\caption{(Color online) 
Band structures for (a) MBN and (c) AB-stacked BLBN are shown across the high-symmetry points in the FBZ. The results are obtained from DFT and compared with the FTB and the simpler effective TB (\(\rm F_2G_2\), \(\rm F_3G_3\) \& \(\rm F_4G_4\)) models with fewer parameters.
(b) For MBN, the FTB \& \(\rm F_4G_4\) models demonstrate superior agreement with { \it ab initio} calculations near the $K$-point and $M$-point compared to \(\rm F_2G_2\) and \(\rm F_3G_3\).
(d) The primary bandgap (\(\rm E_{\rm gap}\)) obtained from DFT for all five stacking types in BLBN is shown by the black line. The direct and indirect bandgap nature is indicated by red and green rectangles, respectively. The dashed blue line represents the energy difference between the primary conduction band minimum at the $M$-point and at the $K$-point for each stacking configuration.} 
\label{fig:3}
\end{figure}

\subsection{Bandstructures }

\begin{figure*}
\centering 
\includegraphics[width=\textwidth]{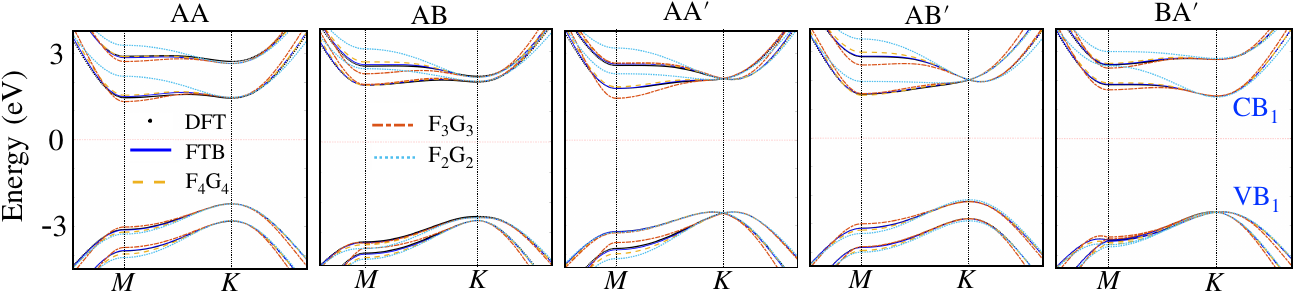}
\caption{(Color online) 
Comparison of the band structures for BLBN stackings (AA, AB, AA$^\prime$, AB$^\prime$, BA$^\prime$), obtained from DFT and compared with the FTB and the simpler effective TB (\(\rm F_2G_2\), \(\rm F_3G_3\) \& \(\rm F_4G_4\)) models. The accuracy of the FTB and \(\rm F_4G_4\) models in reproducing the \textit{ab initio} \(\pi\)-bands along the specified \(\bm{k}\)-path, compared to \(\rm F_2G_2\) and \(\rm F_3G_3\) models which deviate at the \(M\)-point, is shown. The primary valence and conduction bands are labeled as VB$_1$ and CB$_1$, respectively, in the rightmost panel.} 
\label{fig:5}
\end{figure*}
\begin{figure*}
\centering 
\includegraphics[width=18cm]{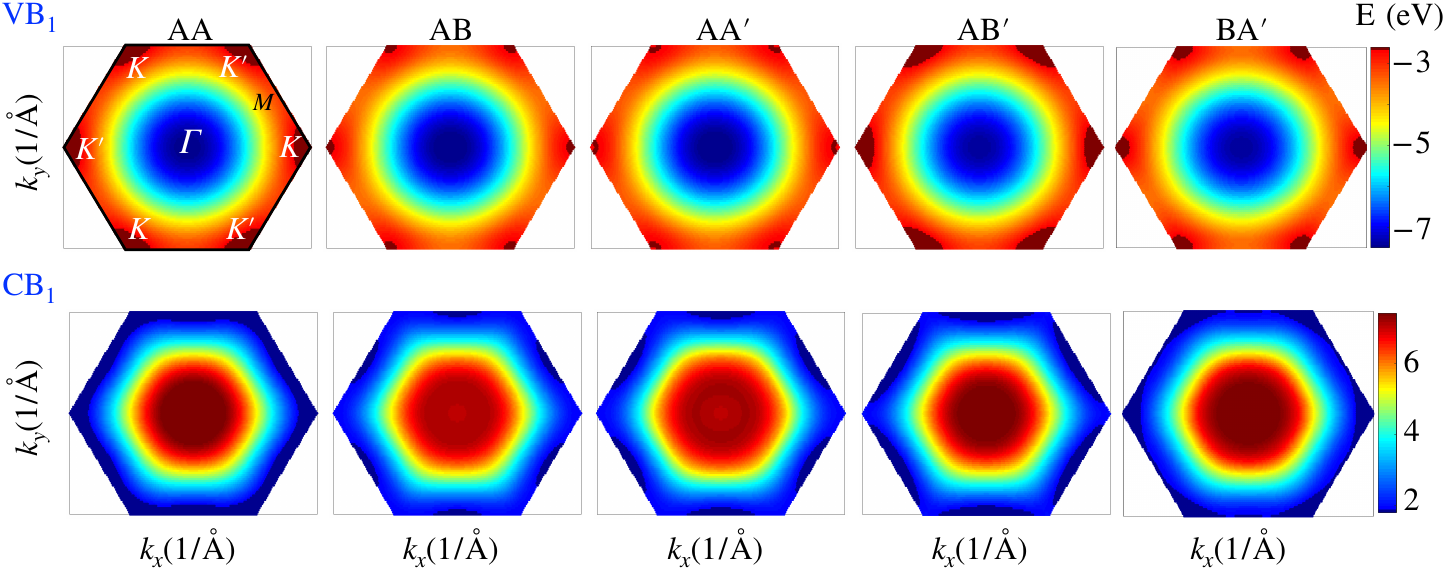}
\caption{(Color online) The surface plots of the primary bands $\rm VB_1$ and $\rm CB_1$, calculated using the FTB model across the entire FBZ for various BLBN stackings, are depicted here. The letters in the top left panel denote the high symmetry points within the FBZ. It is observed that the VB$_1$ maxima occur at the $K$ point (except for AA$^\prime$), while the VB$_1$ minima are consistently found at the $\varGamma$ point. The AA$^\prime$ stacking shows valence band crossings at the $K$ point. The CB$_1$ minima appear at the $M$ point for most stackings but shift to the $K$ point for AA and BA$^\prime$ configurations.
} 
\label{fig:6}
\end{figure*}

We present the electronic band structures calculated for the MBN and BLBN systems using both the FTB and effective models. Fig.~\ref{fig:3} compares the band structures of \textit{h}-BN monolayer and the most stable AB stacked bilayer, obtained from DFT, the FTB model, and simplified effective models incorporating up to \( n = 2, 3, \) and \( 4 \) nearest neighbors. The FTB model accurately reproduces the band edges at the \( K \) and \( M \) points, as well as the overall band behavior across the FBZ. However, for simpler calculations, we developed effective TB models that capture these key band edge features. Starting with the double structure factor model, $\rm F_2G_2$, truncates the G$_n$ and F$_n$ functions at $n=2$. In this model, the nearest neighbor hopping terms are taken from \textit{ab initio} hopping data, while the more distant ($n=2$) terms are corrected using expressions derived from the coefficients in Eqs.~\ref{Eq:cpab0} and \ref{Eq:cab1}, as detailed in previous studies~\cite{previous_study_on_monoG_TBmodel, previous_study_on_BiG_TBmodel}:

\begin{equation}
\begin{aligned}
t_{\alpha\beta2} = C_{\alpha\beta1}/\sqrt3a+t_{\alpha\beta1}/2, \\
t'_{\alpha\beta2} = \frac{1}{6}(C'_{\alpha\beta0}-t'_{\alpha\beta0}+3t'_{\alpha\beta1}   ).    
\end{aligned}
\end{equation}
Here, \( t^{(')}_{\alpha\beta n} \) represents the hopping energy of the \( n^{th} \) nearest neighbor hopping process from sublattice \(\alpha\) to \(\beta\), where the primes indicate expansions of the G$_n$ structure factor terms. For single-layer and bilayer graphene, the $\rm F_{2}G_{2}$ model achieves accuracy near the $K$-point and away from it, capturing both the trigonal distortion of the bands near the $K$-point and the particle-hole symmetry breaking throughout the Brillouin zone \cite{previous_study_on_monoG_TBmodel}. Similarly, for MBN, the $\rm F_{2}G_{2}$ model achieves accuracy near the $K$-point, as shown in the Fig.~\ref{fig:3}(b). Given that MBN exhibits low-energy band edges near $K$, the $\rm F_{2}G_{2}$ model could serve as an alternative to the FTB model. In contrast, for BLBN with different stackings, the lower conduction band edge does not always lie at the $K$-point. In AB, AA$^\prime$, and AB$^\prime$ stackings, the conduction band edges are located at the $M$-point, resulting in an indirect bandgap. Conversely, AA and BA$^\prime$ stackings exhibit a direct bandgap, illustrated in 
Fig.~\ref{fig:5}. The bandgap nature and magnitude in these systems are illustrated in Fig.~\ref{fig:3}(d). The overall energy difference between the primary conduction band edges at the $K$-point and $M$-point for different stackings is approximately $\pm 400$ meV, as indicated by the dashed blue line.
\begin{figure}[htb!]
\begin{center}
\includegraphics[width=6cm]{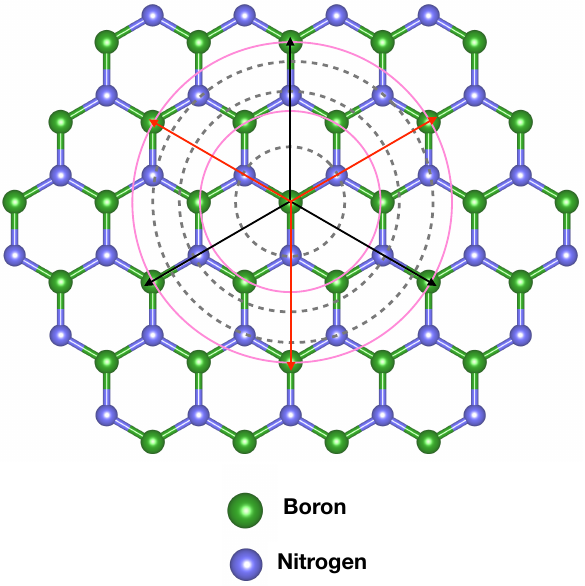}
\caption{(Color online) The nearest neighbor mapping in the TB models is illustrated. Dashed gray circles depict inter sublattice nearest neighbors, while solid pink circles show intra-sublattice nearest neighbors, with the central boron atom at the origin. The index $n$, representing the order of the nearest neighbor, increases with the radius of the circles. 
} 
\label{fig:4}
\end{center}
\end{figure}
Additionally a comprehensive overview is presented in Fig.~\ref{fig:6} showing the surface plots of the lowest valence ($\rm VB_1$) and conduction ($\rm CB_1$) bands across the entire Brillouin zone, calculated using the FTB model. 

While valence band maxima generally occur at the $K$ points (except for AA$^\prime$), valence band minima are consistently found at the $\varGamma$ point. Interestingly, the AA$^\prime$ stacking shows valence band crossings at the $K$ points. Regarding conduction bands, minima appear at the $M$ point for most stackings but shift to the $K$ point for AA and BA$^\prime$ configurations. The $\rm F_{2}G_{2}$ effective model is unable to reproduce these features of bilayers.

As depicted in Fig.\ref{fig:4}, the second coordination shell, represented by the pink large circle, comprises six intra-sublattice (G$_2$) neighbors arranged in an alternating pattern (indicated by black and red arrows). For AB, AA$^\prime$, AB$^\prime$, and BA$^\prime$ stacked bilayers, the interlayer $\rm G_2$ hopping energies between these alternating sublattices differ due to variations in the local atomic environment. The $\rm F_{2}G_{2}$ model, with its limited number of hopping terms and assumption of equal interlayer $\rm G_2$ hopping energies, fails to capture these subtle structural differences, leading to inaccuracies in the calculated band structure near the $M$-point. Therefore, there is a necessity for improved simplified TB models that can accurately capture the complex bandgap behavior of BLBN by effectively reproducing the low-energy bands near both the $K$ and $M$ points, matching the precision of the FTB model.

We further extended these models to include interactions with up to $n = 3$ and $4$ nearest neighbors, denoted as $\rm F_3G_3$ and $\rm F_4G_4$ models, respectively. This inclusion improves the accuracy of the band structure, particularly at regions away from the K-point. In these models, we incorporate \textit{ab initio} hopping data for the shortest hopping terms, while correcting the most distant hopping terms using the following relations:
\\
\\
$~~~~~~~~n=3$:
\begin{equation}
\begin{aligned}
t_{\alpha\beta3} = 2C_{\alpha\beta1}/\sqrt3a+t_{\alpha\beta1}-2t_{\alpha\beta2}, \\ 
t'_{\alpha\beta3} = \frac{1}{3}(-C'_{\alpha\beta0}+t'_{\alpha\beta0}-3t'_{\alpha\beta1}+6t'_{\alpha\beta2}),
\end{aligned}
\end{equation}
\\
$~~~~~~~~n=4$:
\begin{equation}
\begin{aligned}
t_{\alpha\beta4} = \frac{-1}{5}(2C_{\alpha\beta1}/\sqrt3a+t_{\alpha\beta1}-2t_{\alpha\beta2}-t_{\alpha\beta3}), \\ 
t'_{\alpha\beta4} = \frac{1}{6}(-C'_{\alpha\beta0}+t'_{\alpha\beta0}-3t'_{\alpha\beta1}+6t'_{\alpha\beta2}-3t'_{\alpha\beta3})
\end{aligned}
\end{equation}

The effective TB models with $n$ = 2 and 3 neighbors successfully capture the crucial low-energy bands near the $K$-point, including the presence or absence of band crossings for specific stacking configurations. However, their accuracy deviates from {\it ab initio} calculations, particularly away from the K-point. 
Remarkably, the $\rm F_4G_4$ model reproduces the {\it ab initio} low-energy band edges near both the $K$-point and the $M$-point, demonstrating its superior ability to capture the electronic structure across the entire Brillouin zone. This makes the $\rm F_4G_4$ model particularly valuable for detailed electronic structure analysis in \textit{h}-BN systems. The limitations observed in the $\rm F_2G_2$ model are mitigated in higher-order effective models, as evidenced by Tables \ref{tab:I} to \ref{tab:VI}. In these data tables, the four sub-lattices of bilayer \textit{h}-BN, namely $\rm B_1$, $\rm N_1$, $\rm B_2$, and $\rm N_2$, are denoted by the labels $A$, $B$, $A^\prime$, and $B^\prime$, respectively.  Supporting codes and data files used to generate the effective models presented in the tables and figures are accessible at~\cite{BNBN_Draft_files}.

\begin{figure}
\centering
\includegraphics[width=7.5cm]{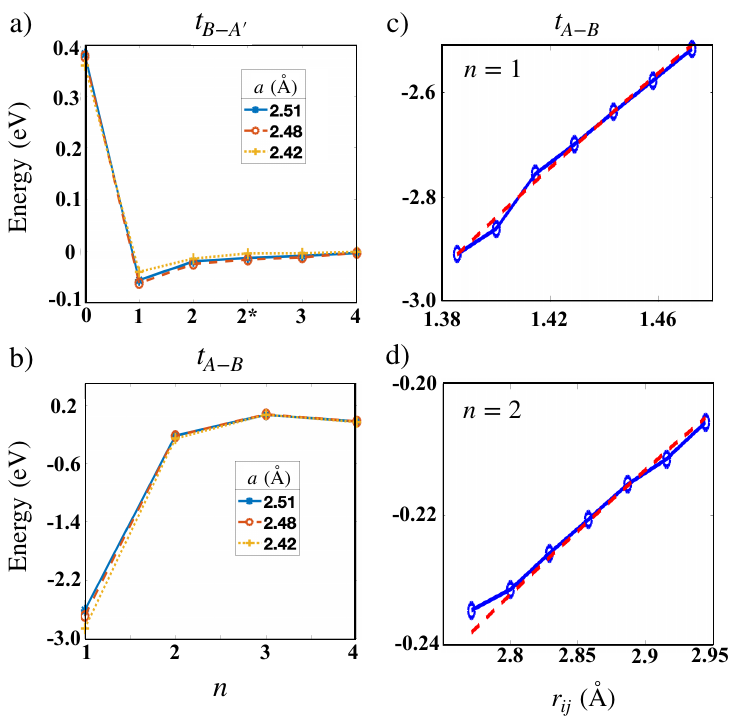}
\caption{(Color online) 
The hopping energies as a function of the nearest-neighbor index (\(n\)) for the AB-stacked BLBN are depicted for two processes: a) interlayer hopping from sublattice B to A$^\prime$ and b) intra-layer hopping from sublattice A to B. Different curves represent results obtained using various lattice constant values. The neighbor index \(n\) signifies the increasing distance between the interacting atoms, with \(n = 2\) and \(2^*\) indicating the interlayer \(\rm G_2\) processes depicted by red and black arrows, respectively, in Fig.~\ref{fig:4}. c) and d) illustrate the intra-layer hopping data \( t_{A-B} \) for nearest neighbors \( n=1 \) and \( n=2 \), respectively, as a function of bond length \( r_{ij} \). The calculated data are shown in blue, while the fitting function, provided in Eq.~(\ref{eq:11}), is depicted in red. }
\label{fig:8}
\end{figure}
\subsection{Effect of strains}

Further, to model the electronic properties of \textit{h}-BN bilayers efficiently using the $\rm F_4G_4$ model, we explored the impact of the lattice constant ($a$) on the effective hopping parameters. Specifically, we computed the hopping terms of the $\rm F_4G_4$ TB model for various lattice constants of the bilayers. It is found that the nearest-neighbor hopping term ($n = 1$) shows significant variation, whereas the more distant ($n = 4$) effective hopping parameters remain largely consistent despite variations in the lattice constants across all hopping processes of each stacking configuration. We have presented the hopping parameters for specific lattice constant values around $a\sim$2.42, 2.48, 2.51~$\rm \AA$ for AB-stacked BLBN in Fig.~\ref{fig:8}(a) \& (b). It is observed that, with a $\sim$1\% variation in $a$, the magnitude of the first nearest neighbor ($n=1$) hopping energy changes by $\sim$3\% in the intra-layer $A$ to $B$ process (corresponding to G$_n$) and by $\sim$2\% in the interlayer $B$ to $A^\prime$ process (corresponding to F$_n$). For understanding the effect of bond distortions on the Fermi velocity in BLBN systems, we have calculated the first nearest neighbor hopping energy for the intra-layer process between $A$-sublattice at site $\bm {r}_i$ to the $B$-sublattice at site $\bm {r}_j$ as a function of bond length, in the most stable AB-stacking. The hopping energy exhibits an exponential dependence on the bond length ($r_{ij} = \left| {\bm r_{ij}} \right| $), which we fit with the following relation: 

\begin{eqnarray}
t_{AB}( r_{ij})=t_{AB}( r_{0,ij})~\exp{(-2.45(\frac{ r_{ij}- r_{0,ij}}{ r_{0,ij}}))}
\label{eq:11}
\end{eqnarray} 
Here, $t_{AB}( r_{0,ij})$ is the hopping energy calculated at the equilibrium bond length $ r_{0,ij} = 1.43~\rm \AA$, which is approximately equal to $-2.7$~eV. There is a good agreement between the hopping data and the fitting function of Eq.~(\ref{eq:11}) as shown in Fig.~\ref{fig:8}(c)\& (d). From the above relation one can account the hopping terms for the lattice constants close to the experimental values.

Additionally, to investigate the dependence of hopping terms on interlayer distances $c$, we computed the $\rm F_{4}G_{4}$ TB model hopping data for all stacking configurations with interlayer distances ranging from $c = 3.1$ to $3.5~ \rm \AA$ in steps of $0.1~ \rm \AA$. We fitted this data with an exponential function of the form:

\begin{equation}
t_{i} (c) = a_i e^{b_{i} c} + c_i e^{d_{i} c}
\label{fitfun}
\end{equation}

Here, $t_{i}(c)$ represents the hopping energy in eV as a function of the interlayer distance. The index $i$ corresponds to the neighbor index $n$ associated with $\rm G$ or $\rm F$ structure factors, ranging up to $n=4$. Specifically, for $\rm G$ structure factor hopping processes, $i$ ranges from 0 to 4, while for $\rm F$ structure factor, $i$ ranges from 1 to 4. This fitting model applies to both intra-layer and interlayer hopping processes between intra- and inter-sublattices across all stackings examined in this study. The corresponding fitting parameters $a_i$, $b_i$, $c_i$, and $d_i$ are detailed in Tables~\ref{tab:A1} to \ref{tab:A5} in Appendix.~\ref{App:Fitting}. In Fig.~\ref{fig:7}(a) and (b), we demonstrate the quality of this fitting for the interlayer ($B$ to $A^\prime$) hopping process and the intra-layer ($A$ to $B$) hopping process in the AB-stacking. 

This fitting function is suitable for all the hopping processes including the distant neighbor terms having the energy of the order of $10^{-3}$~eV. Our fitting parametrization can accurately reproduce the $\rm F_{4}G_{4}$-TB model band structures, at any intermediate interlayer distance. For instance, at the intermediate value of $c$ = 3.261 $\rm \AA$ in the AB-stacking, the fitting function can produce hopping terms as accurate as those listed in Table.~\ref{tab:III}, and the resulting band structures are in close agreement, as compared in Fig.~\ref{fig:7}(c).
The increasing interlayer distance leads to an increase in the energy separation between the primary and secondary bands. Specifically, in Fig.~\ref{fig:7}(d) $\&$ (e), the separation between the primary valence band $\rm VB_1$ and primary conduction band $\rm CB_1$ at the $K$-point is plotted as a function of interlayer distance, obtained from Eq.~\ref{fitfun} for AB-stacking.

\begin{figure}
\centering
\includegraphics[width=8.75cm]{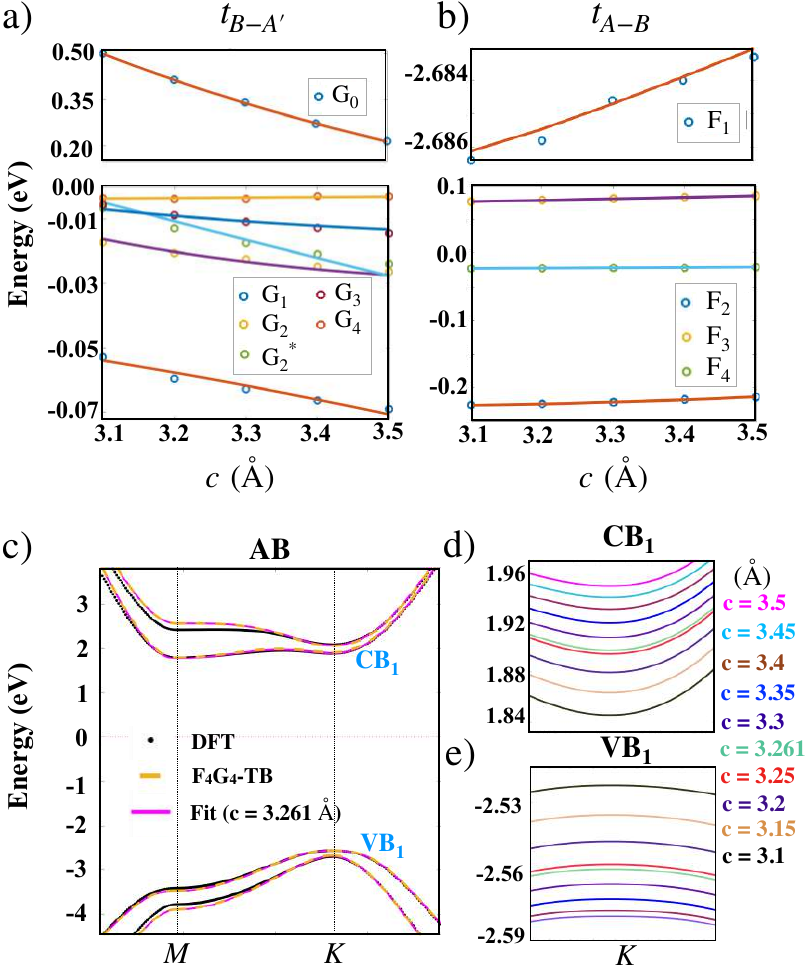}
\caption{(Color online) 
The hopping energy data fitting using Eq.~(\ref{fitfun}) as a function of interlayer distance for AB-stacked BLBN is shown for two processes: a) interlayer hopping between B to A$^\prime$ and b) intra-layer hopping between A to B. Each curve represents different neighbor indices (\(n\)), with the data points denoted by circles. c) Comparison of the band structures for AB-stacked BLBN at \(c = 3.261 \, \text{Å}\), showing results from DFT, the \(\rm F_{4}G_{4}\)-TB model, and the fitted \(\rm F_{4}G_{4}\)-TB model from Eq.~(\ref{fitfun}). At the \(K\)-point, the separation between d) the primary conduction band \(\rm CB_1\) and e) the valence band \(\rm VB_1\)  is illustrated for different values of \(c\) (Å).
}
\label{fig:7}
\end{figure}

\begin{figure*}
\centering
\includegraphics[width=15cm]{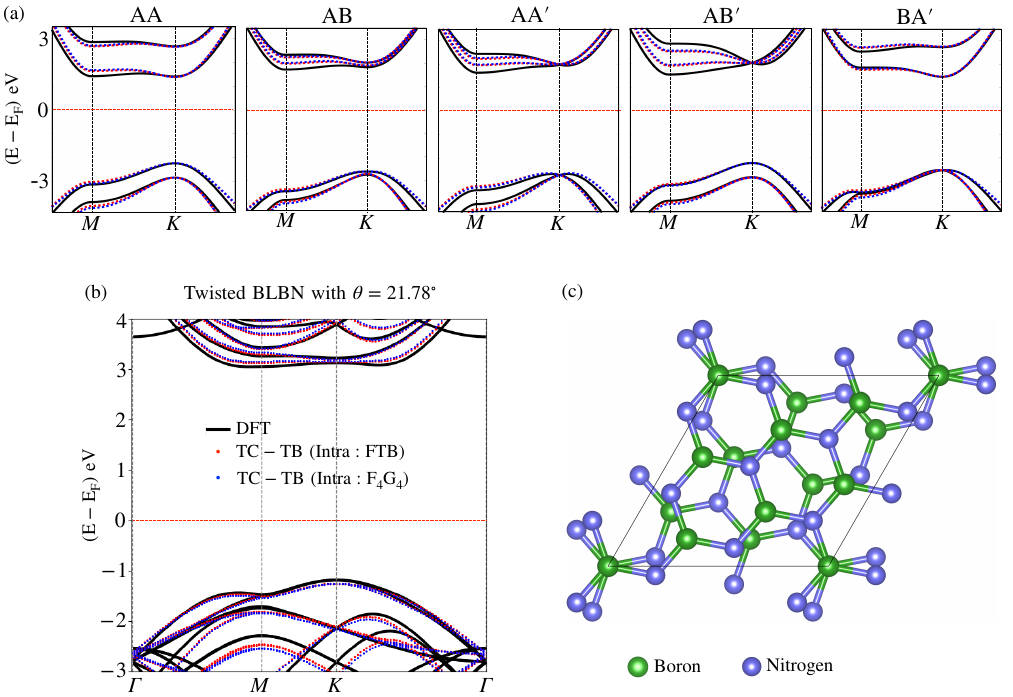}
\caption{(Color online) Comparison of bilayer hBN band structures obtained from DFT and with the TC-approximated TB model, incorporating both FTB and $\rm F_{4}G_{4}$ models for intra-layer terms: (a) across all standard stackings of BLBN, (b) for a twisted BLBN configuration with $\theta = 21.78^\circ$, and (c) the schematic of the corresponding supercell.} 
\label{fig:9}
\end{figure*}
\

\section{Distance-dependent two-center approximation for interlayer hopping processes}

\label{sec:tc-model}

The two-center (TC) approximation offers a generalized TB model approach for studying the electronic structures of twisted {\it h}-BN bilayers with diverse stacking configurations. This approach has proven successful in capturing the electronic properties of twisted bilayer graphene~\cite{rijmayou_1, rijmayou, TCmodel_Koshino, TCmodel_NL}. We present a revised TC approximation that goes beyond fitting DFT bands and incorporates insights from effective models, unlike previous approaches in BLBN~\cite{TCmodel_BLBN1, TCmodel_BLBN2}. Our model builds on the FTB and the well-established $\rm F_4G_4$ model for intra-layer interactions. . 
We incorporate a distance-dependent approach for interlayer hopping, which aligns well with the DFT Hamiltonian at the $K$-point~\cite{jungdxdy}. The distance-dependent interlayer hopping energy in this TC approximation~\cite{rijmayou_1, rijmayou, TCmodel_Koshino, TCmodel_NL} is defined as:

\begin{equation}
\begin{aligned}
t(r_{ij})  = {n}_{ij, z}^2 V_{pp\sigma}(r_{ij}) + (1-{n}_{ij, z}^2) V_{pp\pi}( r_{ij})
\end{aligned}
\label{eq:12}
\end{equation}

Here, $r_{ij}$ is the relative distance between two sublattices positioned at boron or nitrogen atoms from different layers, each located at $\bm {r}_i$ and $\bm {r}_j$, respectively. The $n_{ij, z}$ is the direction cosine of ${r}_{ij}$, defined as ${ n}_{ij, z} = z_{ij}/r_{ij}$, 
where $z_{ij}$ is the coordinate of $ r_{ij}$ along the $z$-axis. 

We have
 
\begin{equation}
\begin{aligned}
V_{pp\sigma}( r_{ij})  = \gamma_1 ~\exp(q_\sigma (1-\frac{ r_{ij}}{c})) , \\
V_{pp\pi}( r_{ij}) = \gamma_0 ~\exp(q_\pi (1-\frac{r_{ij}}{a_{\rm BN}}))
\end{aligned}
\label{eq:13}
\end{equation}
 and 
  \begin{equation}
\begin{aligned}
\frac{q_\sigma}{{c}} = \frac{q_\pi}{{a}_{\rm BN}} = \frac{\ln(\gamma'_0/\gamma_0)}{a_{\rm BN}-a}
\end{aligned}
\label{eq:13_1}
\end{equation}
where we use the nearest neighbor interaction $\gamma_0 = -2.7$~eV within a plane and the second nearest neighbor interaction $\gamma'_0$ as $0.1\gamma_0$~\cite{rijmayou}.
The $a_{\rm BN}$ is the bond length between boron and nitrogen (1.43~$\rm \AA$) and $a$ is the optimized lattice parameter (2.48~$\rm \AA$) within LDA. We consider a constant interlayer distance $c = 3.261~\rm \AA$, to keep the {\it h}-BN layer in the BLBN  to be flat in these calculations. 
The prefactors of the interlayer hopping parameters for boron-boron, nitrogen-nitrogen, and boron-nitrogen interactions within the TC approximation are:
\begin{equation}
\begin{aligned}
\gamma_{1,\,{\rm BB}} = 0.831~eV,  \\
\gamma_{1,\,{\rm NN}} = 0.3989~eV, \\
\gamma_{1,\,{\rm BN}} =  0.6601~eV
\end{aligned}
\label{eq:hoppingterms}
\end{equation}

The $\gamma_1$ terms are obtained by fitting the TB tunneling Hamiltonian element H$({K}: \boldsymbol{d}_{xy})$ with $ab~initio$ data near the $K$-point~\cite{jungdxdy}. Here, $\bm d_{xy}$ represents the registry space of all possible in-plane relative positions for BLBN layers. This space encompasses all stackings achievable by translating one layer relative to the other within the range of $\bm d_x = 0 $ to $a$ and $\bm d_y = 0 $ to $\sqrt3a$. This fitting process aims to minimize the error between the H$({K}: \boldsymbol{d}_{xy})$ data obtained from both the TB model and LDA~\cite{TCmodel_NL}. Substituting these values into Eqs.~(\ref{eq:12}) to (\ref{eq:13_1}), we compute the interlayer hopping energies for various hopping processes within the BLBN system. 

In Fig.~\ref{fig:9} (a), we compare the band structures obtained from LDA calculations with those from the TC-approximated TB model (with intra-layer terms from FTB \& $\rm F_{4}G_{4}$) for various high symmetry stacking configurations of BLBN. The comparison reveals that our TC approximated TB model provides precise band dispersion near the $K$-point, comparable to DFT results. It effectively captures both crossing and non-crossing bands at the $K$-point for all standard stackings. However, it lacks the accuracy away from the $K$-point.

Further, we calculated the band structure for a twisted BLBN supercell with \(\theta = 21.78^\circ\), containing 28 atoms. This supercell was constructed using the optimized lattice constant \(a = 2.48~\AA\) and the equilibrium interlayer distance \(c = 3.261~\AA\) of AB-stacking. In Fig.~\ref{fig:9}(b), we compare the band structures of this twisted BLBN supercell, obtained using both LDA and the TC-approximated TB model, with intra-layer terms defined based on the AA-stacking configuration. Notably, our model shows close agreement with the DFT results. The close match between the bands plotted with red and blue dots, representing the FTB and $\rm F_4G_4$ models used for intra-layer interactions, highlights a significant computational advantage of the $\rm F_4G_4$ model. This advantage is especially pronounced for larger twisted BLBN supercells containing thousands of atoms, indicating that our model is a robust approach for accurately calculating the low-energy bands in such systems.

\section{Summary and conclusions}
\label{sec:summary}
We present the development of $\pi$-band TB models for {\it h}-BN monolayer and bilayers by using maximally localized Wannier functions derived from DFT calculations of the electronic structure. We proposed three effective TB models as simplified alternatives to the complex 15-parameter FTB model. These models focus on accurately describing bands near the high symmetry points ($K$ and $M$) in the FBZ. Our analysis reveals that including hopping terms from up to $n=4$ nearest neighbors (namely the F$_4$G$_4$ model) is necessary to capture the indirect nature of the DFT band gaps observed in various stacking configurations of bilayer {\it h}-BN. The limitation of the simpler $\rm F_{2}G_{2}$ model ($n=2$) stems from its limited number of hopping terms and the assumption of equal interlayer $\rm G_2$ hopping parameters. 
Furthermore, we explore the influence of varying lattice parameters and interlayer distances on the effective hopping parameters in these systems to account for the effect of strains. Through fitting parameterization, we present an interlayer distance-dependent $\rm F_{4}G_{4}$ model for each standard BLBN stacking type.
In addition, we present a TC-approximated TB model that determines interlayer hopping parameters for any stacking or twisted configurations of BLBN. 
Our proposed model integrates the effective $\rm F_{4}G_{4}$ and FTB models for mapping intra-layer interactions, remains robust for calculating low-energy bands in twisted BLBN systems with many atoms per unit cell.

\section{Acknowledgements}
This work was supported by the basic Study and Interdisciplinary R\&D Fund of the University of Seoul (2022).

We acknowledge computational support from KISTI Grant No. KSC-2022-CRE-0514 and by the resources of Urban Big Data and AI Institute (UBAI) at UOS. J.J. also acknowledges support from the Korean Ministry of Land, Infrastructure and Transport (MOLIT) from the Innovative Talent Education Program for Smart Cities.


\begin{appendices}

\appendix

\section{Taylor expansion of the Hamiltonian}
\label{sec:appendixA}

The Taylor expansion of the Hamiltonian near the $K$-point can be expressed as:

\[ H_{\alpha \beta}(\bm{k_D} + \bm{k}) = H_{\alpha \beta}(\bm{k_D}) + \left. \frac{\partial H_{\alpha \beta}}{\partial \bm{k}} \right|_{\bm{k_D}} \bm{k} + \frac{1}{2} \left. \frac{\partial^2 H_{\alpha \beta}}{\partial \bm{k}^2} \right|_{\bm{k_D}} \bm{k}^2 + \cdots \]

For the intra-sublattice processes ($\alpha = \beta$), the first-order term (linear in \(\bm{k}\)) is negligible due to the symmetry operations inherent in the hexagonal lattice. These symmetries lead to a cancellation of contributions when summing over all equivalent positions within the unit cell. 
Therefore, we primarily consider the constant and quadratic terms. We define them as:

\[ C'_{\alpha \beta 0} = H_{\alpha \alpha}(\bm{k_D}) \]
\[ C'_{\alpha \beta 2} = \frac{1}{2} \left. \frac{\partial^2 H_{\alpha \alpha}}{\partial \bm{k}^2} \right|_{\bm{k_D}} \]

Thus, the Hamiltonian for intra-sublattice processes becomes:
\begin{equation}
\begin{aligned}
H_{\alpha \beta}(\bm{k_D}+\bm{k})\simeq C'_{\alpha \beta 0}+C'_{\alpha \beta 2} \bm{k}^2,
\end{aligned}
\label{Eq:Hamil-intra}
\end{equation}
While the off-diagonal elements of the Hamiltonian, corresponding to inter-sublattice ($\alpha \neq \beta$) processes, include both linear and quadratic terms in $\bm k$ with phase factors due to the angular dependence \(\theta_{\bm{k}}\). The zeroth-order term is typically negligible because direct transitions between different sublattices without momentum transfer (\(\bm{k} = 0\)) are significantly reduced in amplitude. Thus, the relevant terms are:

\[ C_{\alpha \beta 1} = \left. \frac{\partial H_{\alpha \beta}}{\partial \bm{k}} \right|_{\bm{k_D}} \]
\[ C_{\alpha \beta 2} = \frac{1}{2} \left. \frac{\partial^2 H_{\alpha \beta}}{\partial \bm{k}^2} \right|_{\bm{k_D}} \]

The Hamiltonian for inter-sublattice processes becomes:
\begin{equation}
\begin{aligned}
 H_{\alpha\beta}(\bm{k_D}+\bm{k})\simeq C_{\alpha\beta1}\bm{k}e^{-i\theta_{\bm{k}}}+C_{\alpha\beta2}\bm{k}^2e^{i2\theta_{\bm{k}}}.
\end{aligned}
\label{Eq:Hamil-inter}
\end{equation}
However, the zeroth-order expansion coefficient for intra-sublattice processes and the first-order expansion coefficient for inter-sublattice processes are essential for constructing the effective Hamiltonian and are derived from the hopping parameters~\cite{previous_study_on_monoG_TBmodel}. They are given by:
\begin{equation}
\begin{aligned}
C'_{\alpha\beta0} =  t'_0-3t'_1+6t'_2-3t'_3-6t'_4+6t'_5+6t'_6-6t'_7 \\
\end{aligned}
\label{Eq:cpab0}
\end{equation}

\begin{equation}
\begin{aligned}
C_{\alpha\beta1} = \frac{\sqrt{3}a}{2} (-t_1+2t_2+t_3-5t_4-4t_5+7t_6+5t_7\\+2t_8-4t_9+11t_{10}) 
\end{aligned}
\label{Eq:cab1}
\end{equation}
Here, $t^{(')}_n$(= $t^{(')}_{\alpha\beta n}$) represents the hopping energy of the $n^{th}$ nearest neighbor hopping process from sublattice $\alpha$ to $\beta$. The primes that are used both for the expansion coefficients and the hopping terms indicate that they involve expansions of G$_n$ structure factor terms. The weights capture the contributions of the $n$-th hopping process. They are are determined by symmetry considerations, including factors like atomic distances, the path orientation of the hopping process, and the phases of the wave functions involved.

\section{ Interlayer distance dependent fitting parametrization tables for $\bf F_4G_4$-TB model in BLBN with stackings}
\label{App:Fitting}

In this section, we provide the data required for the $c$-dependent parametrization of the hopping energies in each hopping process within the effective TB model with $n = 4$ for the BLBN system with different stackings. The hopping data obtained from first-principles, as a function of interlayer distance ranging from 3.1 to 3.5 $\AA$ for the given stackings, is fitted with a function described in Eq.~(\ref{fitfun}). The fitting parameters for all the hopping processes in AA, AA$^\prime$, AB$^\prime$, BA$^\prime$, and AB stackings are listed in Tables \ref{tab:A1}, \ref{tab:A2}, \ref{tab:A3}, \ref{tab:A4}, and \ref{tab:A5} below, respectively.

\setcounter{table}{0}
\renewcommand{\thetable}{B\arabic{table}}
\begin{table}[b]
\begin{tabular}{|lllll|}
\hline
\multicolumn{5}{|c|}{Monolayer $h$-BN}                                                                             \\ \hline
\multicolumn{5}{|c|}{F$_2$G$_2$ model}                                                                               \\ \hline
\multicolumn{1}{|l|}{G$_n$} & \multicolumn{1}{l|}{$t_{AA}$} & \multicolumn{1}{l|}{$t_{BB}$} & \multicolumn{1}{l|}{F$_n$} & $t_{AB}$ \\ \hline
\multicolumn{1}{|l|}{G$_0$} & \multicolumn{1}{l|}{0.1648}    & \multicolumn{1}{l|}{$-$3.8678}    & \multicolumn{1}{l|}{F$_1$} & $-$2.7547    \\ \hline
\multicolumn{1}{|l|}{G$_1$} & \multicolumn{1}{l|}{0.0542}    & \multicolumn{1}{l|}{0.2228}    & \multicolumn{1}{l|}{F$_2$} &$-$0.1329     \\ \hline
\multicolumn{1}{|l|}{G$_2$} & \multicolumn{1}{l|}{0.0566}    & \multicolumn{1}{l|}{0.0429}    & \multicolumn{2}{l|}{}         \\ \hline
\multicolumn{5}{|c|}{F$_3$G$_3$ model}                                                                               \\ \hline
\multicolumn{1}{|l|}{G$_0$} & \multicolumn{1}{l|}{0.1648}    & \multicolumn{1}{l|}{$-$3.8678}    & \multicolumn{1}{l|}{F$_1$} & $-$2.7547    \\ \hline
\multicolumn{1}{|l|}{G$_1$} & \multicolumn{1}{l|}{0.0542}    & \multicolumn{1}{l|}{0.2228}    & \multicolumn{1}{l|}{F$_2$} & $-$0.2362    \\ \hline
\multicolumn{1}{|l|}{G$_2$} & \multicolumn{1}{l|}{0.0397}    & \multicolumn{1}{l|}{0.0329}    & \multicolumn{1}{l|}{F$_3$} &   0.2068  \\ \hline
\multicolumn{1}{|l|}{G$_3$} & \multicolumn{1}{l|}{$-$0.0337}    & \multicolumn{1}{l|}{$-$0.0200}    & \multicolumn{2}{l|}{}         \\ \hline
\multicolumn{5}{|c|}{F$_4$G$_4$ model}                                                                               \\ \hline
\multicolumn{1}{|l|}{G$_0$} & \multicolumn{1}{l|}{0.1648}    & \multicolumn{1}{l|}{$-$3.8678}    & \multicolumn{1}{l|}{F$_1$} &$-$2.7547     \\ \hline
\multicolumn{1}{|l|}{G$_1$} & \multicolumn{1}{l|}{0.0542}    & \multicolumn{1}{l|}{0.2228}    & \multicolumn{1}{l|}{F$_2$} & $-$0.2362    \\ \hline
\multicolumn{1}{|l|}{G$_2$} & \multicolumn{1}{l|}{0.0397}    & \multicolumn{1}{l|}{0.0329}    & \multicolumn{1}{l|}{F$_3$} &0.0539     \\ \hline
\multicolumn{1}{|l|}{G$_3$} & \multicolumn{1}{l|}{$-$0.0361}    & \multicolumn{1}{l|}{$-$0.0250}    & \multicolumn{1}{l|}{F$_4$} &$-$0.0306     \\ \hline
\multicolumn{1}{|l|}{G$_4$} & \multicolumn{1}{l|}{0.0012}    & \multicolumn{1}{l|}{0.0025}    & \multicolumn{2}{l|}{}         \\ \hline
\end{tabular}
\caption{The hopping parameters in eV units for MBN, used to construct the $\rm F_2G_2$, $\rm F_3G_3$, and $\rm F_4G_4$ models, are listed here. The column labels F$_n$ and G$_n$ emphasize that they consist of the hopping terms related to the $f_n$ and $g_n$ structure factors, respectively, for the $n^{th}$ nearest neighbor.}
\label{tab:I}
\end{table}

\begin{table}[t]
\begin{tabular}{|l|l|l|l|l|l|l|l|}
\hline
\multicolumn{8}{|c|}{AA stacked BLBN}                             \\ \hline
\multicolumn{8}{|c|}{F$_2$G$_2$ model}                                         \\ \hline
G$_n$ & $t_{AA}$ &$t_ {BB}$ &  $t_{AA'}$ & $t_{BB'}$ & F$_n$ &$t_ {AB}$ &$t_ {AB'}$  \  \\ \hline
G$_0$ &1.7666    & $-$2.1843     &0.7270       &0.2705              & F$_1$ &$-$2.7001             &0.0265        \\ \hline
G$_1$ &0.0053    &0.1923      &0.0498       & $-$0.0185            & F$_2$ &$-$0.1105              & $-$0.0077       \\ \hline
G$_2$ &0.0471    &0.0370      & 0.0020      &$-$0.0061              & \multicolumn{3}{l|}{}         \\ \hline
\multicolumn{8}{|c|}{F$_3$G$_3$ model}                                         \\ \hline
G$_0$ &1.7666        &$-$2.1843       &0.7270       &0.2705               & F$_1$ & $-$2.7001           &0.0265       \\ \hline
G$_1$ & 0.0053       &0.1923       &  0.0498     & $-$0.0185              & F$_2$ & $-$0.2102          & 0.0082      \\ \hline
G$_2$ & 0.0223       & 0.0195      &  0.0089     & $-$0.0025             & F$_3$ &0.1995             &  $-$0.0317      \\ \hline
G$_3$ & $-$0.0497       & $-$0.0351      & 0.0138    &0.0072              & \multicolumn{3}{l|}{}         \\ \hline
\multicolumn{8}{|c|}{F$_4$G$_4$ model}                                         \\ \hline
G$_0$ &1.7666        &$-$2.1843       &0.7270       & 0.2705              & F$_1$ &$-$2.7001            & 0.0265      \\ \hline
G$_1$ & 0.0053       & 0.1923      &0.0498       &  $-$0.0185             & F$_2$ & $-$0.2102           & 0.0082      \\ \hline
G$_2$ &0.0223        &0.0195       & 0.0089      &   $-$0.0025           & F$_3$ &  0.0797           &  $-$0.0176      \\ \hline
G$_3$ & $-$0.0483       & $-$0.0373      & 0.0139      & 0.0070             & F$_4$ &$-$0.0240           &0.0028        \\ \hline
G$_4$ & $-$0.0007       & 0.0011      & 0.0000       & 0.0001             & \multicolumn{3}{l|}{}         \\ \hline
\end{tabular}
\caption{ For AA-stacked BLBN, the hopping parameters in eV units used to construct the $\rm F_2G_2$, $\rm F_3G_3$, and $\rm F_4G_4$ models are listed here. In the hopping processes of AA-stacking, $t_{A'A'} = t_{AA}$, $t_{B'B'} = t_{BB}$, $t_{A'B'} = t_{AB}$, and $t_{BA'} = t_{AB'}$ by symmetry relations. The column labels F$_n$ and G$_n$ emphasize that they consist of the hopping terms related to the $f_n$ and $g_n$ structure factors, respectively, for the $n^{th}$ nearest neighbor.}
\label{tab:II}
\end{table}

\begin{table*}[t]
\begin{tabular}{|l|l|l|l|l|l|l|l|l|l|l|l|}
\hline
\multicolumn{12}{|c|}{AB  stacked BLBN}                          \\ \hline
\multicolumn{12}{|c|}{F$_2$G$_2$ model}                                      \\ \hline
G$_n$ & $t_{AA}$ &  $t_{BB}$ &  $t_{A'A'}$ &  $t_{B'B'}$ &  $t_{BA'}$ & F$_n$ &  $t_{AB}$ &  $t_{A'B' }$&  $t_{AA'}$ &  $t_{BB'}$& $t_{ AB'}$\  \\ \hline
G$_0$ &1.6636 & $-$2.3393   & 1.7128      &$-$2.2591       &0.3809      & F$_1$ & $-$2.6971    &  $-$2.7190     & 0.4841     &$-$0.0176      & 0.1209      \\ \hline
G$_1$ &0.0235 & 0.1903   &0.0108      & 0.1910       & $-$0.0617     & F$_2 $& $-$0.1248   & $-$0.1129      &0.0457     & $-$0.0545    &  $-$0.1437   \\ \hline
G$_2 $&0.0496  &0.0388     &0.0490       & 0.0328     & $-$0.0158     & \multicolumn{6}{l|}{}            \\ \hline
\multicolumn{12}{|c|}{F$_3$G$_3$ model}                                      \\ \hline
G$_0 $&  1.6636     &$-$2.3393    & 1.7128      &$-$2.2591       & 0.3809     & F$_1$ &$-$2.6971    & $-$2.7190     & 0.4841    &$-$0.0176     &0.1209     \\ \hline
G$_1$ & 0.0235      &0.1903     & 0.0108      &0.1910      &$-$0.0617     & F$_2$ & $-$0.2207   &$-$0.2044      &0.0743     & $-$0.0472    & $-$0.1387    \\ \hline
G$_2$ &0.0274        & 0.0212   & 0.0257     & 0.0148     & $-$0.0245    & F$_3$ &  0.1917   &0.1829      & $-$0.0573    & $-$0.0146    & $-$0.0099    \\ \hline
G$_2^*$ &0.0274        & 0.0212   & 0.0257     & 0.0148     & $-$0.0159   & \multicolumn{6}{l|}{}            \\ \hline
G$_3$ &  $-$0.0446     & $-$0.0352  & $-$0.0466    & $-$0.0359      &  $-$0.0175   & \multicolumn{6}{l|}{}            \\ \hline
\multicolumn{12}{|c|}{F$_4$G$_4$ model}                                      \\ \hline
G$_0 $&1.6636       &$-$2.3393    & 1.7128     & $-$2.2591     & 0.3809    & F$_1$ & $-$2.6971   &$-$2.7190      & 0.4841    &$-$0.0176     &0.1209     \\ \hline
G$_1$ &0.0235       &0.1903    &0.0108      &0.1910      & $-$0.0617    & F$_2$ &$-$0.2207    & $-$0.2044     & 0.0743    & $-$0.0472    &$-$0.1387     \\ \hline
G$_2$ & 0.0274      & 0.0212   &0.0257      & 0.0148     &$-$0.0245     & F$_3$ & 0.0779   & 0.0795     &0.0520     &0.0207     & $-$0.0215    \\ \hline
G$_2^*$ & 0.0274      & 0.0212   &0.0257      & 0.0148     & $-$0.0159    & F$_4$ & $-$0.0228   & $-$0.0207     & 0.0219    &0.0071     &$-$0.0023     \\ \hline
G$_3$ &$-$0.0419       &$-$0.0367    & $-$0.0441     & $-$0.0372     & $-$0.0116    &\multicolumn{6}{l|}{}            \\ \hline
G$_4$ &$-$0.0013       &0.0007    &$-$0.0012      &0.0007      & $-$0.0030    & \multicolumn{6}{l|}{}            \\ \hline
\end{tabular}
\caption{
For AB-stacked BLBN, the hopping parameters in eV units used to construct the $\rm F_2G_2$, $\rm F_3G_3$, and $\rm F_4G_4$ models are listed here. The column labels F$_n$ and G$_n$ emphasize that they consist of the hopping terms related to the $f_n$ and $g_n$ structure factors, respectively, for the $n^{th}$ nearest neighbor.}
\label{tab:III}
\end{table*}

\begin{table*}[t]
\begin{tabular}{|l|l|l|l|l|l|l|l|l|l|}
\hline
\multicolumn{10}{|c|}{AA$^\prime$ stacked BLBN}                             \\ \hline
\multicolumn{10}{|c|}{F$_2$G$_2$ model}                                         \\ \hline
G$_n$ & $t_{AA}$ &$t_ {BB}$ & $t_{A'A'}$ & $t_{B'B'}$ & $t_{AA'}$  & F$_n$ &$t_ {AB}$ &$t_ {AB'}$ & $t_{BA'}$ \  \\ \hline
G$_0$ &1.6717    & $-$2.3068     &$-$2.3074       &1.6716       &0.4310            & F$_1$ &  $-$2.7049    &  0.4239      & $-$0.0613       \\ \hline
G$_1$ & 0.0122   &0.1900      &0.1900       &0.0122       & $-$0.0684           & F$_2$ & $-$0.1176     &0.1338       &$-$0.0161        \\ \hline
G$_2$ &0.0520    &0.0372      & 0.0372      & 0.0520      &$-$0.0344            & \multicolumn{4}{l|}{}         \\ \hline
\multicolumn{10}{|c|}{F$_3$G$_3$ model}                                         \\ \hline
G$_0$ & 1.6717       &  $-$2.3068     & $-$2.3074      &1.6716       &0.4310            & F$_1$ &  $-$2.7049   & 0.4239     & $-$0.0613      \\ \hline
G$_1$ &0.0122        & 0.1900      &0.1900       & 0.0122      &  $-$0.0684          & F$_2$ &  $-$0.2136    &0.1798      &  0.0031     \\ \hline
G$_2$ &0.0288        &0.0186       & 0.0186      & 0.0288      & $-$0.0415          & F$_3$ &  0.1921    &$-$0.0919      & $-$0.0383       \\ \hline
G$_2^*$ &0.0288        &0.0186       & 0.0186      & 0.0288      & $-$0.0163          & \multicolumn{4}{l|}{}         \\ \hline
G$_3$ &$-$0.0463        &$-$0.0370       &  $-$0.0370   &$-$0.0463       &$-$0.0142            & \multicolumn{4}{l|}{}         \\ \hline
\multicolumn{10}{|c|}{F$_4$G$_4$ model}                                         \\ \hline
G$_0$ &1.6717        &$-$2.3068       &$-$2.3074       & 1.6716      & 0.4310           & F$_1$ & $-$2.7049    & 0.4239            &   $-$0.0613    \\ \hline
G$_1$ & 0.0122       & 0.1900      & 0.1900      &0.0122       &  $-$0.0684           & F$_2$ &$-$0.2136     & 0.1798            & 0.0031      \\ \hline
G$_2$ &0.0288        &0.0186       & 0.0186      & 0.0288      &$-$0.0415     & F$_3$ &0.0803      &   0.0261        &0.0121        \\ \hline
G$_2^*$ & 0.0288       &0.0186       &0.0186       & 0.0288      & $-$0.0163      & F$_4$ &$-$0.0224      &0.0236             &   0.0101     \\ \hline
G$_3$ &$-$0.0438        &$-$0.0381       &  $-$0.0381     &$-$0.0438       &$-$0.0072        &  \multicolumn{4}{l|}{}         \\ \hline
G$_4$ & $-$0.0013       &0.0005       &  0.0005     & $-$0.0013      &$-$0.0035        & \multicolumn{4}{l|}{}         \\ \hline
\end{tabular}
\caption{For AA$^\prime$-stacked BLBN, the hopping parameters in eV units used to construct the $\rm F_2G_2$, $\rm F_3G_3$, and $\rm F_4G_4$ models are listed here. In the hopping processes of AA$^\prime$-stacking, $t_{BB'}$ = $t_{AA'}$ and $t_{A'B'}$= $t_{AB}$ by symmetry relations. The column labels F$_n$ and G$_n$ emphasize that they consist of the hopping terms related to the $f_n$ and $g_n$ structure factors, respectively, for the $n^{th}$ nearest neighbor.}
\label{tab:IV}
\end{table*}

\begin{table}[t]
\begin{tabular}{|l|l|l|l|l|l|l|l|}
\hline
\multicolumn{8}{|c|}{AB$^\prime$ stacked BLBN}                          \\ \hline
\multicolumn{8}{|c|}{F$_2$G$_2$ model}                                      \\ \hline
G$_n$ & $t_{AA}$ &  $t_{BB}$ &  $t_{BA'}$ & F$_n$ &  $t_{AB}$ &  $t_{AA'}$ &   $t_{ AB'}$\  \\ \hline
G$_0$ &1.8176        & $-$2.1740          &0.2059      & F$_1$ &$-$2.6892     &    0.0687    &  0.4460      \\ \hline
G$_1$ & 0.0183       & 0.1911           &$-$0.0283      & F$_2 $& $-$0.1183    & $-$0.0823    &  $-$0.0686     \\ \hline
G$_2 $& 0.0451       &0.0377          &0.0129      & \multicolumn{4}{l|}{}            \\ \hline
\multicolumn{8}{|c|}{F$_3$G$_3$ model}                                      \\ \hline
G$_0 $& 1.8176      & $-$2.1740        & 0.2059     & F$_1$ & $-$2.6892        & 0.0687    &  0.4460   \\ \hline
G$_1$ & 0.0183      &0.1911          & $-$0.0283    & F$_2$ & $-$0.2104   & $-$0.0771         &$-$0.0328     \\ \hline
G$_2$ & 0.0219       &0.0205         & 0.0170    & F$_3$ & 0.1842   &  $-$0.0103        &  $-$0.0717   \\ \hline
G$_2^*$ & 0.0219       &0.0205         &  $-$0.0064   &  \multicolumn{4}{l|}{}            \\ \hline
G$_3$ & $-$0.0464      &$-$0.0343          & 0.0083    & \multicolumn{4}{l|}{}            \\ \hline
\multicolumn{8}{|c|}{F$_4$G$_4$ model}                                      \\ \hline
G$_0 $&1.8176       &  $-$2.1740  & 0.2059          & F$_1$ &$-$2.6892    &0.0687           &0.4460     \\ \hline
G$_1$ &0.0183       & 0.1911   &  $-$0.0283         & F$_2$ &  $-$0.2104  &$-$0.0771           & $-$0.0328    \\ \hline
G$_2$ &0.0219       & 0.0205   & 0.0170          & F$_3$ &0.0752    &   $-$0.0054       &  0.0452   \\ \hline
G$_2^*$ &0.0219       & 0.0205   & $-$0.0064        & F$_4$ &$-$0.0218    &   0.0010       &  0.0234   \\ \hline
G$_3$ &$-$0.0431       &$-$0.0356    &0.0063           &  \multicolumn{4}{l|}{}            \\ \hline
G$_4$ &$-$0.0016       & 0.0007   &0.0010           & \multicolumn{4}{l|}{}            \\ \hline
\end{tabular}
\caption{For AB$^\prime$-stacked BLBN, the hopping parameters in eV units used to construct the $\rm F_2G_2$, $\rm F_3G_3$, and $\rm F_4G_4$ TB models are listed here. In the hopping processes of AB$^\prime$-stacking, $t_{A'A'}$ = $t_{BB}$, $t_{B'B'}$ = $t_{AA}$, $t_{A'B'}$ = $t_{AB}$, and $t_{BB'}$ = $t_{AA'}$ by symmetry relations. The column labels F$_n$ and G$_n$ emphasize that they consist of the hopping terms related to the $f_n$ and $g_n$ structure factors, respectively, for the $n^{th}$ nearest neighbor.}
\label{tab:V}
\end{table}

\begin{table}[t]
\begin{tabular}{|l|l|l|l|l|l|l|l|}
\hline
\multicolumn{8}{|c|}{BA$^\prime$stacked BLBN}                          \\ \hline
\multicolumn{8}{|c|}{F$_2$G$_2$ model}                                      \\ \hline
G$_n$ & $t_{AA}$ &  $t_{BB}$  &  $t_{AB'}$ & F$_n$ &  $t_{AB}$ &   $t_{AA'}$ & $t_{ BA'}$\  \\ \hline
G$_0$ &  1.8325      &$-$2.1688         &  0.7813    & F$_1$ & $-$2.6866    & 0.0763     &   0.0604      \\ \hline
G$_1$ & 0.0096       & 0.1965          & 0.0877     & F$_2 $&$-$0.1202     & $-$0.0569    &  $-$0.0735    \\ \hline
G$_2 $& 0.0497       &0.0379          &  0.0242    & \multicolumn{4}{l|}{}            \\ \hline
\multicolumn{8}{|c|}{F$_3$G$_3$ model}                                      \\ \hline
G$_0 $& 1.8325      & $-$2.1688   &0.7813             & F$_1$ &  $-$2.6866  & 0.0762          & 0.0604    \\ \hline
G$_1$ & 0.0096        & 0.1965    &  0.0877         & F$_2$ & $-$0.2187   & $-$0.0622        &$-$0.0665     \\ \hline
G$_2$ & 0.0244        &0.0204     &   0.0312       & F$_3$ & 0.1971   &     0.0106      & $-$0.0140    \\ \hline
G$_2^*$ & 0.0244        &0.0204     & 0.0193       &  \multicolumn{4}{l|}{}            \\ \hline
G$_3$ &$-$0.0507        &$-$0.0352   &0.0140           & \multicolumn{4}{l|}{}            \\ \hline
\multicolumn{8}{|c|}{F$_4$G$_4$ model}                                      \\ \hline
G$_0 $& 1.8325      &  $-$2.1688  & 0.7813          & F$_1$ &  $-$2.6866  & 0.0763         & 0.0604    \\ \hline
G$_1$ & 0.0096      & 0.1965   &   0.0877        & F$_2$ & $-$0.2187   &$-$0.0622           &$-$0.0665     \\ \hline
G$_2$ & 0.0244      & 0.0204   & 0.0312          & F$_3$ &  0.0794   &   $-$0.0141       & 0.0122    \\ \hline
G$_2^*$ & 0.0244      & 0.0204   &  0.0193          & F$_4$ & $-$0.0235   &   $-$0.0049       & 0.0052    \\ \hline
G$_3$ &$-$0.0466       &  $-$0.0370  &0.0093          & \multicolumn{4}{l|}{}            \\ \hline
G$_4$ &  $-$0.0020     & 0.0009   &   0.0024        & \multicolumn{4}{l|}{}            \\ \hline
\end{tabular}

\caption{For BA$^\prime$-stacked BLBN, the hopping parameters in eV units used to construct the $\rm F_2G_2$, $\rm F_3G_3$, and $\rm F_4G_4$ TB models are listed here. In the hopping processes of BA$^\prime$-stacking, $t_{A'A'}$ = $t_{BB}$, $t_{B'B'}$ = $t_{AA}$, $t_{A'B' }$ = $t_{AB}$, and $t_{BB'}$ = $t_{AA'}$ by symmetry relations. The column labels F$_n$ and G$_n$ emphasize that they consist of the hopping terms related to the $f_n$ and $g_n$ structure factors, respectively, for the $n^{th}$ nearest neighbor.}
\label{tab:VI}
\end{table}

\begin{table}[b]
\begin{tabular}{|l|l|l|l|l|l|l|}
\hline
   & ${A-A}$ & ${B-B} $& ${A-A^\prime}$ & ${B-B^\prime}$ & ${A-B}$ & ${A-B^\prime}$ \\ \hline
$a_0$ & 0.4606  & $-$4.0140  & 2.2170  & 1.6800  &  -      &     -       \\ \hline
$b_0 $& 0.3915  & $-$0.1848  & $-$2.7200 & $-$0.1375           &  -       & -           \\ \hline
$c_0$ & 0.0256  & $-$0.0024  & 1.4480  & $-$0.2480  &  -       &   -         \\ \hline
$d_0$ & 0.4038  &  $-$4.6280 & $-$0.2130  & 0.3563  &  -       &     -       \\ \hline
   &           &           &            &            &           &            \\ \hline
$a_1$ & $-$0.0093 & 0.6383   & $-$0.3910       & $-$0.1087   &$-$0.9425         &1.9030           \\ \hline
$b_1$ & 0.6518  & 0.1833  & 0.6001       & 0.1701    & $-$8.8350         & $-$0.9914          \\ \hline
$c_1$ & 0.0055  &  $-$0.4867   & 0.3238       &0.0715   & $-$2.9180  &  $-$0.0007         \\ \hline
$d_1$ & 0.8256  &  0.2112  & 0.6629       &0.2693     & $-$0.0238  & 1.2810          \\ \hline
   &           &           &            &            &           &            \\ \hline
$a_2$ & $-$0.0002 & 0.0268   & 0.0202       &0.0095   & $-$0.1598  & 8.8100           \\ \hline
$b_2$ & 0.6784 & 0.4281     & 0.7543       &0.3499 & 0.0820 &   $-$2.8300         \\ \hline
$c_2$ & 0.0040 & $-$0.0191   &$-$0.0403   & $-$0.0157   & 9.2690   & 1.3430         \\ \hline
$d_2$ & 0.5384 & 0.4734    & 0.5324       & 0.2181   & $-$7.2540 &  $-$1.5260        \\ \hline
   &           &           &            &            &           &            \\ \hline
$a_3$ & 0.5653   & 0.0108  & 0.1714      & 0.4662   &0.0015 & $-$0.1925          \\ \hline
$b_3$ & $-$0.2976 & 0.5824   & $-$0.1436 &0.7221& $-$1.2870 &  $-$0.0979          \\ \hline
$c_3$ & $-$0.6604 & $-$0.0367  & $-$0.3223  &$-$0.4688  & 0.0782         &  0.4838          \\ \hline
$d_3$ & $-$0.2825 & 0.3377 & $-$0.3789&  0.7200  &  0.0111        & $-$0.4207          \\ \hline
   &           &           &            &            &           &            \\ \hline
$a_4$ & $-$0.0077 & $-$0.0005 & $-$0.0725  & 0.0002  &$-$0.0454          &  $-$0.1698          \\ \hline
$b_4$ & 0.5426  & $-$9.3350 &$-$0.6360  & 0.8318   &  $-$0.1396        &  0.1638          \\ \hline
$c_4$ & 0.0072 & 0.0010   & 0.0065       & $-$0.0006  & 0.0214         & 0.2067            \\ \hline
$d_4$ & 0.5580  & 0.0058 & 0.0999       &  0.4771  & $-$0.2948         &  0.1068         \\ \hline
\end{tabular}
\caption{The fitting parameters $a_i$, $b_i$, $c_i$, and $d_i$ for the $c$-dependent hopping parametrization shown in Eq.~(\ref{fitfun}), for the $\rm F_4G_4$ model for AA-stacked BLBN are presented. Here, $i$ represents the neighbor index $n$. The top row of the table illustrates the hopping processes between corresponding sublattices. In AA-stacking, $t_{A'A'}$ = $t_{AA}$, $t_{B'B'}$ = $t_{BB}$, $t_{A'B'}$ = $t_{AB}$, and $t_{BA'}$ = $t_{AB'}$ by symmetry relations. The hopping parameters are given in eV units. 
}
\label{tab:A1}
\end{table}

\begin{table}[t]
\begin{tabular}{|l|l|l|l|l|l|l|}
\hline
   & ${A-A}$ & ${B-B} $& ${A-A^\prime}$  & ${A-B}$ & ${A-B'}$& ${B-A'}$ \\ \hline
$a_0$ &0.0164 & 0.4819 &2.8020 &   -        &     -  &     -    \\ \hline
$b_0 $&$-$3.0860 &0.2315 &$-$0.3861 & -          & -      & -     \\ \hline
$c_0$ &1.5040 &$-$1.7470&$-$0.0073 &  -         & -      & -    \\ \hline
$d_0$ &0.0303&0.1979 &1.1910 &  -         &    -   &     -    \\ \hline
   &           &           &            &                     &       &     \\ \hline
$a_1$ & $-$0.0931 &0.3269 & $-$0.0279 &$-$2.4920 & $-$5.4770 &$-$0.8401 \\ \hline
$b_1$ & $-$0.2723 &$-$0.1689& 0.2853&$-$9.2210 &$-$6.0280 & 0.2899\\ \hline
$c_1$ &11.750 & $-$1.7430 &7.1910 &$-$2.6770  &1.8530 &0.6791 \\ \hline
$d_1$ &$-$1.6680  &$-$7.4570 & $-$9.0300 &0.0038&$-$0.4450 &0.3461 \\ \hline
   &           &           &            &                        &        &    \\ \hline
$a_2$ &0.3006& 3.9530 & 0.9573 & 0.0056 & $-$0.2072 & 1.0550 \\ \hline
$b_2$ & 0.3349 & $-$1.0680 &0.1134& $-$0.0310 &$-$0.3630  &0.1763 \\ \hline
$c_2$ & $-$0.2654 &$-$4.3010&$-$0.9907 &$-$0.4010&0.6907 &$-$1.1360 \\ \hline
$d_2$ &0.3630 &$-$1.1420 &0.1122 &$-$0.1920&$-$0.3184 & 0.1531\\ \hline
   &           &           &            &                        &        &    \\ \hline
$a_2^*$&0.3006& 3.9530 &0.2170 &  -         &   -   &    -\\ \hline
$b_2^*$&0.3349 &$-$1.0680 &0.4179 &     -      &  -    &  - \\ \hline
$c_2^*$ &$-$0.2654 &$-$4.3010 &$-$0.2164 &     -      &   -   & -   \\ \hline
$d_2^*$ &0.3630 &$-$1.1420  &0.4246 &   -        &   -   &    - \\ \hline
   &           &           &            &                        &        &    \\ \hline
$a_3$ & $-$0.0187&$-$0.0178 & $-$0.0703&0.0285& 0.1277&0.9009\\ \hline
$b_3$ &0.2587 &0.2275 &0.3113 &0.3129 & 0.2866 &0.5398 \\ \hline
$c_3$ & $-$0.0838 &$-$0.6783 &0.0801& 0.0061  &  $-$0.1503 & $-$0.9120\\ \hline
$d_3$ &$-$1.8550& $-$9.6650  & 0.2584&$-$10.520  & 0.2081 &0.5353 \\ \hline
   &           &           &            &                        &        &    \\ \hline
$a_4$ &0.1765 &0.0204 &0.1122&0.0183 & 0.0147 &0.0028 \\ \hline
$b_4$ &0.3853 &$-$1.1480 &$-$0.1774 &$-$9.8800 &0.0422& 0.3726 \\ \hline
$c_4$ &$-$0.1565 &0.1483 &$-$0.1163 &$-$0.0536 & 0.0046 &$-$3.7130 \\ \hline
$d_4$ &0.4226&$-$3.4070 &$-$0.1702 &$-$0.2787 &0.1945 &$-$4.746\\ \hline
\end{tabular}

\caption{The fitting parameters $a_i$, $b_i$, $c_i$, and $d_i$ for the $c$-dependent hopping parametrization shown in Eq.~(\ref{fitfun}), for the $\rm F_4G_4$ model for AA$^\prime$-stacked BLBN are presented. Here, $i$ represents the neighbor index $n$. The top row of the table illustrates the hopping processes between corresponding sublattices. In AA$^\prime$-stacking, $t_{A'A'} = t_{BB}$, $t_{B'B'} = t_{AA}$, $t_{A'B'} = t_{AB}$, and $t_{BB'} = t_{AA'}$ by symmetry relations. The hopping parameters are given in eV units.
}
\label{tab:A2}
\end{table}

\begin{table}[t]
\begin{tabular}{|l|l|l|l|l|l|l|}
\hline
   & ${A-A}$ & ${B-B} $& ${B-A^\prime}$  & ${A-B}$ & ${A-A'}$& ${A-B'}$ \\ \hline
$a_0$ & 0.9895&$-$0.1395 &0.8377&   -      &     -     &     -    \\ \hline
$b_0 $&0.2307&$-$4.9030 & $-$0.0721& -        & -         & -     \\ \hline
$c_0$ &$-$0.0034&$-$2.2510 &$-$0.0643 &  -       & -         & -    \\ \hline
$d_0$ &1.3550 &$-$0.0105 & 0.5943& -        &     -     &     -   \\ \hline
   &           &           &            &                     &       &     \\ \hline
$a_1$ &4.1740 &1.2190 &$-$0.2888 & 1.8790 & 3.9740  & $-$0.0145 \\ \hline
$b_1$ &$-$1.6680 &$-$6.9440&$-$5.5960 & $-$9.1540 &$-$0.7872  & 0.6848 \\ \hline
$c_1$ &0.1304&0.2440 &$-$10.570 &$-$2.7100 &$-$0.4307 & 1.0160 \\ \hline
$d_1$ &$-$4.5090 &$-$0.0753&$-$1.8350 &$-$0.0023  &$-$0.1839 & $-$0.1714\\ \hline
   &           &           &            &                        &        &    \\ \hline
$a_2$ &$-$0.0037&$-$0.0112 & 0.0840 & $-$0.3066 &$-$0.0613 &0.0646 \\ \hline
$b_2$ &0.7808&0.7136 &$-$9.7680& $-$0.1052 & 0.0712 & 0.1191 \\ \hline
$c_2$ &0.0104&0.0235 &0.0011& 8.1740  & $-$2.8940 &$-$14.240 \\ \hline
$d_2$ &0.5810 &0.5375 &0.8206 &$-$2.1490  &$-$7.2330  &$-$1.4450 \\ \hline
   &           &           &            &                        &        &    \\ \hline
$a_2^*$ &$-$0.0037&$-$0.0112 &$-$0.5739 &     -    &   -       &  -  \\ \hline
$b_2^*$ &0.7808&0.7136 & 0.3430 &   -      &  -        &  - \\ \hline
$c_2^*$ &0.0104&0.0235 &0.5383 &  -       &     -     &  - \\ \hline
$d_2^*$ &0.5810&0.5375 &0.3615&        - &   -       &   -  \\ \hline
   &           &           &            &                        &        &    \\ \hline
$a_3$ &0.0469 & 7.2940 & $-$0.0343 &0.0312 &5.2910  & $-$1.3750  \\ \hline
$b_3$ &$-$4.1440 &$-$9.0010&0.6859 &0.2687 &$-$1.7530 & 0.5064 \\ \hline
$c_3$ &$-$0.0225& $-$0.0157&0.0310 & 0.0000&$-$0.0022 & 1.3750 \\ \hline
$d_3$ &0.1979&0.2507 &0.7234 & $-$10.900 &0.7159  & 0.5083  \\ \hline
   &           &           &            &                        &        &    \\ \hline
$a_4$ &0.2959&0.0045 &0.0065 & $-$0.0324 & 0.4313& $-$0.0016 \\ \hline
$b_4$ &$-$1.0540& $-$0.2546&-0.2343 &$-$0.1242 &0.2359& 0.6225 \\ \hline
$c_4$ &$-$0.1077&$-$0.0004&-2.2330&2.5360 &$-$0.3990 & 0.0171 \\ \hline
$d_4$ &$-$0.6937&0.3998 &-2.1300 &$-$12.150  &0.2594  &0.2214 \\ \hline
\end{tabular}

\caption{
The fitting parameters $a_i$, $b_i$, $c_i$, and $d_i$ for the $c$-dependent hopping parametrization shown in Eq.~(\ref{fitfun}), for the $\rm F_4G_4$ model for AB$^\prime$-stacked BLBN are presented. Here, $i$ represents the neighbor index $n$. The top row of the table illustrates the hopping processes between corresponding sublattices. In AB$^\prime$-stacking, $t_{A'A'} = t_{BB}$, $t_{B'B'} = t_{AA}$, $t_{A'B'} = t_{AB}$, and $t_{BB'} = t_{AA'}$ by symmetry relations. The hopping parameters are given in eV units.}
\label{tab:A3}
\end{table}

\begin{table}[]
\begin{tabular}{|l|l|l|l|l|l|l|}
\hline
   & ${A-A}$ & ${B-B} $& ${A-B^\prime }$  & ${A-B}$ & ${A-A'}$& ${B-A'}$ \\ \hline
$a_0$ &1.7450 &$-$0.4610&$-$1.0390&   -      &     -    &     -   \\ \hline
$b_0 $&0.4394&0.4089&0.7784&-         & -        &-    \\ \hline
$c_0$ &$-$0.8599&$-$60.580 &1.3910& -        &-         & -   \\ \hline
$d_0$ &0.5683&$-$1.5240 &0.7067 & -        &     -    &    -   \\ \hline
   &           &           &            &                     &       &     \\ \hline
$a_1$ &0.2493& 0.0759 &$-$0.0556 &$-$2.6490  &3.6570 & 8.0410 \\ \hline
$b_1$ &0.4731&0.1196&1.1320& 0.0043 & $-$0.7636 &$-$0.1605 \\ \hline
$c_1$ &$-$0.2362& 0.2223&0.0465 &0.3333  &$-$0.1874 &$-$7.0360  \\ \hline
$d_1$ &0.4874&$-$0.3001&1.1980 &$-$2.4060   & 0.05863  &$-$0.1236  \\ \hline
   &           &           &            &                        &        &    \\ \hline
$a_2$ &0.0634&0.3877 &0.0002& $-$0.2814 &0.1185 &  5.5470 \\ \hline
$b_2$ &0.3311&$-$0.1194 & 1.4740 &$-$0.0789 & $-$9.8480 & $-$2.3650 \\ \hline
$c_2$ &$-$0.0482&$-$0.3550 & 0.0000 &0.0000 &$-$0.0280&$-$3.0960 \\ \hline
$d_2$ &0.3728&$-$0.1166 &$-$9.9250 & $-$15.660 & 0.2427  &$-$1.1660 \\ \hline
   &           &           &            &                        &        &    \\ \hline
$a_2^*$ &0.0634&0.3877 &$-$0.0779 & -       & -        &  -    \\ \hline
$b_2^*$ &0.3311&$-$0.1194 & 0.4572 &      -  &    -     &  -   \\ \hline
$c_2^*$ &$-$0.0482&$-$0.3550 &0.0682 &   -     &     -    & - \\ \hline
$d_2^*$ &0.3728&$-$0.1166 &0.5146&    -    &   -      &  -   \\ \hline
   &           &           &            &                        &        &    \\ \hline
$a_3$ &$-$0.1957&0.0062 &$-$0.3733 &  $-$1.7680 &0.4363  &  0.0000     \\ \hline
$b_3$ &$-$0.6550&0.1607 &$-$0.4989& $-$7.6770 & $-$0.4018 &$-$14.260\\ \hline
$c_3$ &$-$0.0027&$-$0.0306 &0.1213 & 0.0595&$-$0.1385 & 0.0002 \\ \hline
$d_3$ &0.6550&0.1345 & $-$0.1180 & 0.0894 &$-$0.0170 &1.2130 \\ \hline
   &           &           &            &                        &        &    \\ \hline
$a_4$ &$-$0.0725&0.1227 &0.0550 & 2.2160&$-$0.0660 &0.4512 \\ \hline
$b_4$ &$-$1.1070&$-$1.5420& 0.4468 & $-$10.750 &0.4113 &$-$0.0064 \\ \hline
$c_4$ &0.9846&0.0000     &$-$0.0619&$-$0.0282 & 0.0908  &$-$0.4692\\ \hline
$d_4$ &$-$4.0320&$-$13.8600 & 0.4074& $-$0.0593 &0.3079& $-$0.0221\\ \hline
\end{tabular}

\caption{
The fitting parameters $a_i$, $b_i$, $c_i$, and $d_i$ for the $c$-dependent hopping parametrization shown in Eq.~(\ref{fitfun}), for the $\rm F_4G_4$ model for BA$^\prime$-stacked BLBN are presented. Here, $i$ represents the neighbor index $n$. The top row of the table illustrates the hopping processes between corresponding sublattices. In BA$^\prime$-stacking, $t_{A'A'} = t_{BB}$, $t_{B'B'} = t_{AA}$, $t_{A'B'} = t_{AB}$, and $t_{BB'} = t_{AA'}$ by symmetry relations. The hopping parameters are given in eV units.
}
\label{tab:A4}
\end{table}

\begin{table*}[]
\begin{tabular}{|l|l|l|l|l|l|l|l|l|l|l|}
\hline
   & ${A-A}$ & ${B-B}$ & ${A'-A'}$ & ${B'-B'}$ & ${B-A'}$ & ${A-B}$ & ${A'-B'}$ & ${A-A'}$ & ${B-B'}$ & ${A-B'}$ \\ \hline
$a_0$ &1.4020  &0.0108 &2.5140 &0.0783 &$-$0.0400 &      -     &    -         &     -      &     -       &    -       \\ \hline
$b_0$ &0.0526  &0.6491 &$-$6.3220 &0.9919 &0.3796 &    -       &   -          &    -      &    -      &  -        \\ \hline
$c_0$ &-0.0988 & $-$2.5820 &3.6710 & $-$0.7260 &42.640 &      -     &          -   &   -         &        -  &    -      \\ \hline
$d_0$ & $-$9.7670 & $-$0.0221 &$-$0.2225 & 0.5373&$-$1.359 &       -    &      -       &       -     &    -        &       -     \\ \hline
   &           &           &            &            &            &           &             &            &            &            \\ \hline
$a_1$ & 1.4180 &$-$0.0553 &$-$0.1604 &0.3536 &$-$0.0059 & $-$2.8050 &$-$0.0222 &$-$0.5973& $-$0.1576 & $-$0.0386 \\ \hline
$b_1$ & $-$0.0319 &0.1875 &0.0707 &$-$0.1884 &0.7141 &$-$0.0033 &$-$4.3200  & 0.9163& $-$0.3161 &0.3746 \\ \hline
$c_1$ & $-$1.2310 &0.3135 & 0.2551& 0.8459 &4.2250 & 0.0924 &$-$2.6880  & 0.7005&  1.3570 & 9.9740   \\ \hline
$d_1$ & 0.0069 &$-$0.0232 &$-$0.0568 &$-$4.8530 &$-$8.9810 &$-$0.0101 & 0.0012  &0.8792 &$-$1.0680   &$-$1.1130  \\ \hline
   &           &           &            &            &            &           &             &            &            &            \\ \hline
$a_2$ &0.5976 &0.1805 & $-$0.1176 & $-$0.0004 &$-$0.5243 &$-$2.8900  & 0.7870 & $-$0.6613 &0.0008 & $-$0.2680  \\ \hline
$b_2$ &$-$0.8391 &$-$0.6548 &0.3172 & 1.4970 &$-$0.5917 &$-$7.4550   & $-$0.2730 & 0.8112 &0.8209  & $-$0.2156   \\ \hline
$c_2$ & $-$54.380 &$-$0.0378 &0.1424 &0.0025 &5.0960 &$-$0.3349 &$-$0.9677 &0.6552& $-$0.1054& $-$0.9209  \\ \hline
$d_2$ & $-$2.5620&$-$1.4540 &0.2810 &1.0030&$-$1.3950 &$-$0.1262 &$-$0.1840  &0.8161&$-$0.1697 &$-$6.4990  \\ \hline
   &           &           &            &            &            &           &             &            &            &            \\ \hline
$a_2^*$ &0.5976  &0.1805 & $-$0.1176 &$-$0.0004 & 0.4423 &-& -  & -& -&            \\ \hline
$b_2^*$ &$-$0.8391 &$-$0.6548&0.3172 &1.4970 &$-$0.2127 &- & - &- & -&            \\ \hline
$c_2^*$ & $-$54.3800 &$-$0.0378 &0.1424 &0.0025 & $-$0.2044 &- & - & -& -&             \\ \hline
$d_2^*$ & $-$2.5620 & $-$1.4540 &0.2810 &1.0030 &0.0444 &- &- &- &-&           \\ \hline
   &           &           &            &            &            &           &             &            &            &            \\ \hline
$a_3$ &0.0000  &$-$0.0798 &0.0000  & $-$0.2003 &$-$0.7780 & 0.0381 &0.0364& $-$0.0089&0.1372 &$-$0.5453  \\ \hline
$b_3$ &$-$18.650 &$-$0.8135& $-$13.680  & $-$0.9921&$-$0.6396 &0.2253 &0.2405&0.6663 &0.2613 & $-$0.2354 \\ \hline
$c_3$ &$-$0.0207 &$-$0.0073 &$-$0.0273 &$-$0.0036 &1.6600 &$-$1.9910 &$-$4.233 &0.0128& $-$0.1557 & 1.4150  \\ \hline
$d_3$ & 0.2181  &0.4485 & 0.1478 &0.6417 &$-$0.9052& $-$7.3900 &$-$5.304 &0.7016 &0.2034 & $-$0.5496 \\ \hline
   &           &           &            &            &            &           &             &            &            &            \\ \hline   
$a_4 $& $-$4.3320 &0.2453 & $-$4.0280 &0.0421 &0.0155 &0.2825  & $-$0.5826 &0.0097 &0.0026 & $-$0.6390   \\ \hline
$b_4$ & $-$7.4290  &$-$1.4510&$-$3.7300 &0.1121 & $-$10.720 &0.0066 &0.4629 & 0.2387 & 0.3089 & $-$2.7570  \\ \hline
$c_4$ &$-$0.0005& $-$0.0968 &$-$0.0003 &$-$0.0377 &$-$0.0132 & $-$0.3180 &0.5664 & 0.7201& $-$1.8140 &  4.0990 \\ \hline
$e_4$ &0.3698 &$-$1.2520 & 0.4917 &0.1420 &$-$0.4014 & $-$0.0066 &0.4690  &$-$3.8410 &$-$5.5240 & $-$3.9340  \\ \hline

\end{tabular}

\caption{
The fitting parameters $a_i$, $b_i$, $c_i$, and $d_i$ for the $c$-dependent hopping parametrization shown in Eq.~(\ref{fitfun}), for the $\rm F_4G_4$ model for AB-stacked BLBN are presented. Here, $i$ represents the neighbor index $n$. The top row of the table illustrates the hopping processes between corresponding sublattices. The hopping parameters are given in eV units.
}
\label{tab:A5}
\end{table*}

\end{appendices}

\end{document}